\input harvmac.tex

\input epsf.tex
\def\figin{\epsfcheck\figin}\def\figins{\epsfcheck\figins}
\def\epsfcheck{\ifx\epsfbox\UnDeFiNeD
\message{(NO epsf.tex, FIGURES WILL BE IGNORED)}
\gdef\figin##1{\vskip2in}\gdef\figins##1{\hskip.5in}% blank space instead
\else\message{(FIGURES WILL BE INCLUDED)}%
\gdef\figin##1{##1}\gdef\figins##1{##1}\fi}
\def\DefWarn#1{}
\def\figinsert{\goodbreak\midinsert}
\def\ifig#1#2#3{\DefWarn#1\xdef#1{fig.~\the\figno}
\writedef{#1\leftbracket fig.\noexpand~\the\figno}%
\figinsert\figin{\centerline{#3}}\medskip\centerline{\vbox{\baselineskip12pt
\advance\hsize by -1truein\noindent\footnotefont{\bf Fig.~\the\figno:} #2}}
\bigskip\endinsert\global\advance\figno by1}
%%%%%%%%%%%%%%%%%%%%%%%%%%%%%%%%%%%%%%%%%%%%%%%%%%%%%%%%%%%%%%%%%%%%%%%%%%%%%

% \draftmode

%\draft

\def\cd{D\!\!\!\!/}

\def\sp{\partial\!\!\!\!/}

\lref\WittYau{E.~Witten and S.~T.~Yau, ``Connectedness of the
boundary in the AdS/CFT correspondence,'' Adv.\ Theor.\ Math.\
Phys.\  {\bf 3}, 1635 (1999) [arXiv:hep-th/9910245].
%%CITATION = HEP-TH 9910245;%%
}

\lref\Romans{M.~Gunaydin, L.~J.~Romans and N.~P.~Warner, ``Gauged
N=8 Supergravity In Five-Dimensions,'' Phys.\ Lett.\ B {\bf 154},
268 (1985).
 %%CITATION = PHLTA,B154,268;%%
}

\lref\Gates{S.~J.~Gates and B.~Zwiebach, ``Gauged N=4 Supergravity The
ory With A New Scalar Potential,'' Phys.\
Lett.\ B {\bf 123}, 200 (1983).}

\lref\Das{A.~Das, ``SO(4) Invariant Extended Supergravity,'' Phys.\ Re
v.\ D {\bf 15}, 2805 (1977).}

\lref\Pope{H.~Lu, C.~N.~Pope and T.~A.~Tran, ``Five-dimensional N
= 4, SU(2) x U(1) gauged supergravity from type IIB,'' Phys.\
Lett.\ B {\bf 475}, 261 (2000) [arXiv:hep-th/9909203].
%%CITATION = HEP-TH 9909203;%%
}

\lref\Awada{ M.~Awada and P.~K.~Townsend, ``D = 8
Maxwell-Einstein Supergravity,'' Phys.\ Lett.\ B {\bf 156}, 51
(1985).
%%CITATION = PHLTA,B156,51;%%
}

\lref\SeibWit{N.~Seiberg and E.~Witten, ``The D1/D5 system and
singular CFT,'' JHEP {\bf 9904}, 017 (1999)
[arXiv:hep-th/9903224].
%%CITATION = HEP-TH 9903224;%%
}

\lref\MalNuna{J.~M.~Maldacena and C.~Nunez, ``Towards the large N
limit of pure N = 1 super Yang Mills,'' Phys.\ Rev.\ Lett.\  {\bf
86}, 588 (2001) [arXiv:hep-th/0008001].
%%CITATION = HEP-TH 0008001;%%
}

\lref\MalNunb{J.~M.~Maldacena and C.~Nunez, ``Supergravity
description of field theories on curved manifolds and a no  go
theorem,'' Int.\ J.\ Mod.\ Phys.\ A {\bf 16}, 822 (2001)
[arXiv:hep-th/0007018].
%%CITATION = HEP-TH 0007018;%%
}

\lref\jpms{ J.~Polchinski and M.~J.~Strassler, ``The string dual of a confining four-dimensional gauge theory,''
arXiv:hep-th/0003136.
%%CITATION = HEP-TH 0003136;%%
}

\lref\MalBTZ{J.~M.~Maldacena, ``Eternal black holes in
Anti-de-Sitter,'' arXiv:hep-th/0106112.
 %%CITATION = HEP-TH 0106112;%%
}

\lref\Tsikas{T.~T.~Tsikas, ``Consistent Truncations Of Chiral N=2
D = 10 Supergravity On The Round Five Sphere,'' Class.\ Quant.\
Grav.\  {\bf 3}, 733 (1986).
%%CITATION = CQGRD,3,733;%%
}

\lref\Vafa{C.~Vafa and E.~Witten, ``A Strong coupling test of S
duality,'' Nucl.\ Phys.\ B {\bf 431}, 3 (1994)
[arXiv:hep-th/9408074].
%%CITATION = HEP-TH 9408074;%%
}

\lref\sgas{ S.~B.~Giddings and A.~Strominger,
``Axion Induced Topology Change In Quantum Gravity And String Theory,''
Nucl.\ Phys.\ B {\bf 306}, 890 (1988).
%%CITATION = NUPHA,B306,890;%%
}

\lref\Cvetic{M.~Cvetic {\it et al.}, ``Embedding AdS black holes in te
n and eleven dimensions,'' Nucl.\ Phys.\ B
{\bf 558}, 96 (1999) [arXiv:hep-th/9903214].
%%CITATION = HEP-TH 9903214;%%
}

\lref\Chamblin{A.~Chamblin, R.~Emparan, C.~V.~Johnson and R.~C.~Myers,
 ``Charged AdS black holes and catastrophic
holography,'' Phys.\ Rev.\ D {\bf 60}, 064018 (1999) [arXiv:hep-th/990
2170].
%%CITATION = HEP-TH 9902170;%%
}

\lref\Shenker{P.~Kraus, H.~Ooguri, S.~Shenker, ``Inside the Horizon wi
th AdS/CFT,'' [arXiv:hep-th/0212277].}

\lref\Sseven{M.~Cvetic, H.~Lu and C.~N.~Pope, ``Four-dimensional N = 4
, SO(4) gauged supergravity from D = 11,''
Nucl.\ Phys.\ B {\bf 574}, 761 (2000) [arXiv:hep-th/9910252].}

\lref\StSusy{Y.~Park, M.~Srednicki and A.~Strominger, ``Wormhole
Induced Supersymmetry Breaking In String Theory,'' Phys.\ Lett.\ B
{\bf 244}, 393 (1990). }

\lref\adstwo{M.~Astorino, S.~Cacciatori, D.~Klemm and D.~Zanon,
``AdS(2) supergravity and superconformal quantum mechanics,''
Annals Phys.\  {\bf 304}, 128 (2003) [arXiv:hep-th/0212096]. }

\lref\Porrati{L.~Girardello, M.~Petrini, M.~Porrati and
A.~Zaffaroni, ``The supergravity dual of N = 1 super Yang-Mills
theory,'' Nucl.\ Phys.\ B {\bf 569}, 451 (2000)
[arXiv:hep-th/9909047]. }

\lref\Herrings{S.~R.~Coleman, ``Black Holes As Red Herrings: Topologic
al Fluctuations And The Loss Of Quantum
Coherence,'' Nucl.\ Phys.\ B {\bf 307}, 867 (1988).}
%%CITATION = NUPHA,B307,867;%%

\lref\Freedman{D.~Z.~Freedman, S.~D.~Mathur, A.~Matusis and L.~Rastelli, ``Correlation functions in the
CFT($d$)/AdS($d+1$) correspondence,'' Nucl.\ Phys.\ B {\bf 546}, 96 (1999) [arXiv:hep-th/9804058].
%%CITATION = HEP-TH 9804058;%%
}

\lref\jm{ J.~M.~Maldacena, ``The large N limit of superconformal field theories and supergravity,'' Adv.\ Theor.\
Math.\ Phys.\  {\bf 2}, 231 (1998) [Int.\ J.\ Theor.\ Phys.\  {\bf 38}, 1113 (1999)] [arXiv:hep-th/9711200].
%%CITATION = HEP-TH 9711200;%%
} \lref\Wittenhol{ E.~Witten, ``Anti-de Sitter space and holography,'' Adv.\ Theor.\ Math.\ Phys.\  {\bf 2}, 253
(1998) [arXiv:hep-th/9802150].
%%CITATION = HEP-TH 9802150;%%
} \lref\gkp{ S.~S.~Gubser, I.~R.~Klebanov and A.~M.~Polyakov, ``Gauge theory correlators from non-critical string
theory,'' Phys.\ Lett.\ B {\bf 428}, 105 (1998) [arXiv:hep-th/9802109].
%%CITATION = HEP-TH 9802109;%%
}

\lref\GHP{ G.~W.~Gibbons, S.~W.~Hawking and M.~J.~Perry, ``Path Integrals And The Indefiniteness
Of The Gravitational Action,'' Nucl.\ Phys.\ B {\bf 138}, 141 (1978).}

\lref\susykill{J.~P.~Gauntlett, J.~B.~Gutowski, C.~M.~Hull, S.~Pakis and H.~S.~Reall, ``All supersymmetric
solutions of minimal supergravity in five dimensions,'' Class.\ Quant.\ Grav.\  {\bf 20}, 4587 (2003)
[arXiv:hep-th/0209114] .\ J.~P.~Gauntlett and S.~Pakis, ``The geometry of D = 11 Killing spinors '' JHEP {\bf
0304}, 039 (2003) [arXiv:hep-th/0212008].\  J.~P.~Gauntlett and J.~B.~Gutowski, ``All supersymmetric solutions
of minimal gauged supergravity in five  dimensions,'' Phys.\ Rev.\ D {\bf 68}, 105009 (2003)
[arXiv:hep-th/0304064] .\ J.~B.~Gutowski, D.~Martelli and H.~S.~Reall, ``All supersymmetric solutions of
minimal supergravity in six dimensions,'' Class.\ Quant.\ Grav.\  {\bf 20}, 5049 (2003) [arXiv:hep-th/0306235].\
 M.~M.~Caldarelli and D.~Klemm, ``All supersymmetric solutions of N = 2, D = 4 gauged supergravity,'' JHEP
{\bf 0309}, 019 (2003) [arXiv:hep-th/0307022].\  M.~M.~Caldarelli and D.~Klemm, ``Supersymmetric Goedel-type
universe in four dimensions,'' arXiv:hep-th/0310081.}

\lref\PvN{P.~van Nieuwenhuizen and A.~Waldron, ``On Euclidean spinors and Wick rotations,'' Phys.\ Lett.\ B {\bf
389}, 29 (1996) [arXiv:hep-th/9608174].}

\lref\WittenWYseq{M.~l.~Cai and G.~J.~Galloway,
``Boundaries of zero scalar curvature in the AdS/CFT correspondence,''
Adv.\ Theor.\ Math.\ Phys.\  {\bf 3}, 1769 (1999)
[arXiv:hep-th/0003046].}
%%CITATION = HEP-TH 0003046;%%

\lref\WHsts{S.~B.~Giddings and A.~Strominger,
``String Wormholes,''Phys.\ Lett.\ B {\bf 230}, 46 (1989).\  S.~Pratik Khastgir and J.~Maharana,
``Topology changing processes and symmetries of string effective action,''
Nucl.\ Phys.\ B {\bf 406}, 145 (1993)
[arXiv:hep-th/9302069].\  G.~W.~Gibbons, M.~B.~Green and M.~J.~Perry,
``Instantons and Seven-Branes in Type IIB Superstring Theory,''
Phys.\ Lett.\ B {\bf 370}, 37 (1996)
[arXiv:hep-th/9511080].\  J.~Y.~Kim, H.~W.~Lee and Y.~S.~Myung,
``D-instanton and D-wormhole,''
Phys.\ Lett.\ B {\bf 400}, 32 (1997)
[arXiv:hep-th/9612249].\
 J.~Y.~Kim, H.~W.~Lee and Y.~S.~Myung,
``Negative modes in the four-dimensional stringy wormholes,''
Phys.\ Rev.\ D {\bf 56}, 6684 (1997)
[arXiv:hep-th/9701116].\  M.~Gutperle and M.~Spalinski,
``Supergravity instantons and the universal hypermultiplet,''
JHEP {\bf 0006}, 037 (2000)
[arXiv:hep-th/0005068].}

\lref\WHnonSts{   M.~Cadoni and M.~Cavaglia, ``Cosmological and
wormhole solutions in low-energy effective string theory,'' Phys.\
Rev.\ D {\bf 50}, 6435 (1994) [arXiv:hep-th/9406053].\
V.~D.~Dzhunushaliev, ``5D Kaluza - Klein's Wormhole Between Two
Event Horizons,'' arXiv:gr-qc/9603007.\  H.~Kim, ``Axionic
wormholes: More on their classical and quantum aspects,'' Nucl.\
Phys.\ B {\bf 527}, 311 (1998) [arXiv:hep-th/9706015].\
V.~D.~Dzhunushaliev and H.~J.~Schmidt, ``Wormholes and flux tubes
in the 7D gravity on the principal bundle with  SU(2) gauge group
as the extra dimensions,'' Phys.\ Rev.\ D {\bf 62}, 044035 (2000)
[arXiv:gr-qc/9911080].\  S.~A.~Hayward, S.~W.~Kim and
H.~j.~Lee, ``Dilatonic wormholes: Construction, operation,
maintenance and collapse  to black holes,'' Phys.\ Rev.\ D {\bf
65}, 064003 (2002) [arXiv:gr-qc/0110080].\  S.~I.~Vacaru and
D.~Singleton, ``Ellipsoidal, cylindrical, bipolar and toroidal
wormholes in 5D gravity,'' J.\ Math.\ Phys.\  {\bf 43}, 2486
(2002) [arXiv:hep-th/0110272].\  M.~Gutperle and W.~Sabra,
``Instantons and wormholes in Minkowski and (A)dS spaces,'' Nucl.\
Phys.\ B {\bf 647}, 344 (2002) [arXiv:hep-th/0206153].\
K.~A.~Bronnikov and S.~W.~Kim, ``Possible wormholes in a brane
world,'' Phys.\ Rev.\ D {\bf 67}, 064027 (2003)
[arXiv:gr-qc/0212112]. }

\lref\Hosoya{A.~Hosoya and W.~Ogura,
``Wormhole Instanton Solution In The Einstein Yang-Mills System,''
Phys.\ Lett.\ B {\bf 225}, 117 (1989).}

\lref\Donets{E.~E.~Donets and D.~V.~Galtsov,
``Continuous Family Of Einstein Yang-Mills Wormholes,''
Phys.\ Lett.\ B {\bf 294}, 44 (1992)
[arXiv:gr-qc/9209008].}

\lref\ColemanTZ{
S.~R.~Coleman and K.~M.~Lee,
``Big Wormholes And Little Interactions,''
Nucl.\ Phys.\ B {\bf 341}, 101 (1990).
%%CITATION = NUPHA,B341,101;%%
}

\lref\ColemanLeeB{S.~R.~Coleman and K.~M.~Lee,
``Wormholes Made Without Massless Matter Fields,''
Nucl.\ Phys.\ B {\bf 329}, 387 (1990).}

\lref\ColemanAlpha{S.~R.~Coleman,
``Why There Is Nothing Rather Than Something: A Theory Of The Cosmological Constant,''
Nucl.\ Phys.\ B {\bf 310}, 643 (1988).\  S.~R.~Coleman and K.~M.~Lee,
``Escape From The Menace Of The Giant Wormholes,''
Phys.\ Lett.\ B {\bf 221}, 242 (1989).}

\lref\ReyAxion{S.~J.~Rey,
``The Axion Dynamics In Wormhole Background,''
Phys.\ Rev.\ D {\bf 39}, 3185 (1989).
}

\lref\Bers{L.~Bers, ``Simultaneous Uniformization,'' Bull.\ Amer.\ Math.\ Soc.\ {\bf 66},
94 (1960).\ L.~Bers, ``Spaces of Kleinian groups,'' In Maryland conference in
several complex variables I, p. 9-34, Springer-Verlag lecture notes in Math. No. 155, 1970.}

\lref\eternal{J.~M.~Maldacena,
``Eternal black holes in Anti-de-Sitter,''
JHEP {\bf 0304}, 021 (2003)
[arXiv:hep-th/0106112].
}

\lref\shenker{P.~Kraus, H.~Ooguri and S.~Shenker,
``Inside the horizon with AdS/CFT,''
Phys.\ Rev.\ D {\bf 67}, 124022 (2003)
[arXiv:hep-th/0212277].
}

\lref\asadjoan{
V.~Balasubramanian, A.~Naqvi and J.~Simon,
``A multi-boundary AdS orbifold and DLCQ holography: A universal holographic description of extremal black hole horizons,''
arXiv:hep-th/0311237.}

\lref\mathHyp{ R.~R.~Mazzeo, ``Unique continuation of infinity and embedded eigenvalues for asymptotically
hyperbolic manifolds,'' Am.\ J.\ Math.\ {\bf 113} (1991), 25-45.\ R.~Schoen and S.~T.~Yau, `` Conformally flat
manifolds, Kleinian groups and scalar curvature,'' J.\ Differential Geometry {\bf 20} (1984), 479-495.\ J.~M.~Lee,
``The spectrum of an asymptotically hyperbolic Einstein manifold,'' Comm.\ Anal.\ Geom.\ {\bf 3} (1995), no. 1-2,
253--271. [arXiv:dg-ga/9409003]. }
\lref\sjrcomp{S.~J.~Rey, ``The Collective Dynamics And The
Correlations Of Wormholes In Quantum Gravity,'' Nucl.\ Phys.\ B {\bf 319}, 765 (1989).\ S.~J.~Rey, ``The Axion
Dynamics In Wormhole Background,'' Phys.\ Rev.\ D {\bf 39}, 3185 (1989).}
%%%%%%%%%%%%%%%%%%%%%%%%%%%%%%%%%%%%%%%%%%%%%%%%%%%%%%%%%%%%%%%%%%

{ \Title{\vbox{\baselineskip12pt \hbox{hep-th/0401024 }  {\vbox{\baselineskip12pt \hbox{ITFA-2003-57 }}} }}
 {\vbox{
%{\centerline { Multiple boundary gravity solutions  }}
%{\centerline{ and some comments on their Field Theory duals }} }}}
{\centerline { Wormholes in AdS  }} }}}
\bigskip
\centerline{ Juan Maldacena$^{1}$ and Liat Maoz$^{2}$ }
\bigskip
\centerline{$^1$ Institute for Advanced Study, Princeton, NJ
08540,USA}

\bigskip
\centerline{$^2$ Institute for Theoretical Physics, University of Amsterdam,} \centerline{Valckenierstraat 65,
1018XE Amsterdam, The Netherlands}

%\bigskip
%\centerline{$^3$ Jefferson Physical Laboratory, Cambridge, MA 02138, USA}

%\bigskip
%\centerline{$^4$ Jadwin Hall, Princeton, NJ 08544,USA}

\vskip .3in

We construct a few Euclidean supergravity solutions  with multiple boundaries. We
consider examples where the corresponding boundary field theory is well defined on each boundary. We point out
that these configurations are puzzling from the AdS/CFT point of view. A proper understanding of the AdS/CFT
dictionary for these cases might yield some information about the physics of closed universes.

\vfill

% \Date{December  2002}
\eject

\newsec{Introduction}

The general statement of the AdS/CFT correspondence is thought
to be that the sum over all geometries with fixed boundary conditions
is the same as the partition function of a
 (conformal) field theory living on the boundary \refs{\jm,\Wittenhol,\gkp}.

A puzzle arises if we have Euclidean
geometries that have more than one disconnected
boundary and are connected through the bulk\foot{
Lorentzian geometries with multiple asymptotic boundaries
 are such that
boundaries are separated by horizons \ref\GallowayBR{
G.~J.~Galloway, K.~Schleich, D.~Witt and E.~Woolgar,
``The AdS/CFT correspondence conjecture and topological censorship,''
Phys.\ Lett.\ B {\bf 505}, 255 (2001)
[arXiv:hep-th/9912119].
%%CITATION = HEP-TH 9912119;%%
}, as long as the boundaries have more than one dimension.
 In this case one can interpret the geometries as dual to
 entangled states of the various field theories living
on each boundary.}. In this paper we explore this puzzle. Our main results are the construction and analysis of a
variety of examples of  Euclidean geometries with two boundaries. We have found examples where the field theory at
each of the boundaries seems perfectly well defined. The  theorem
in \refs{\WittYau,\WittenWYseq}, shows that one cannot find examples
with positive boundary curvature which solve  Einstein's equations
with a negative cosmological constant. We found examples once we
turn on
additional Yang-Mills fields.

The configurations we describe are similar to wormholes in the
sense that they connect two well understood asymptotic regions.
Our geometries are solutions of the ten or eleven dimensional
supergravity actions that arise in string theory. In previous
examples of wormhole-like solutions either the two asymptotic
regions were not as well understood or they were not solutions
of the ten or eleven dimensional supergravity actions
\refs{\sgas,\WHsts,\WHnonSts,\ColemanLeeB,\ColemanTZ,\Hosoya,\Donets,\sjrcomp}

In section two we discuss some general properties of the solutions
we have found. In section three we discuss the simplest examples based
on quotients of hyperbolic space. In section four we discuss an
example closely related to the standard $AdS_4 \times S^7$ solution, where
the boundary field theory is twisted by the fields of a meron.
In section five we discuss an example closely related to the
$AdS_5 \times S^5$ solution, where the boundary theory is twisted by
an instanton.  In section six we point out some of the puzzles
raised by these solutions and we speculate on their possible resolution.

\newsec{ Generalities}

In this paper we consider only Euclidean solutions. We will
analyze solutions which are    asymptotically AdS. We study
solutions that have two disconnected boundaries, which are
connected through the interior. In this sense they are similar to
the Euclidean wormholes considered in \sgas\
 which connect two asymptotically flat regions.
The typical form of the metric for these solutions is \eqn\metric{
ds^2 = d \rho^2 +  w(\rho)^2 d s^2_{\Sigma_{d}} } where $\Sigma$
is a compact surface and $w(\rho) \sim e^{ |\rho|} $ as $ \rho \to
\pm \infty$. The two disconnected boundaries are at $\rho = \pm
\infty$. Most solutions we consider are reflection symmetric
around $\rho =0$. These solutions have the interesting property
that they  can  be analytically continued  into Lorentzian
signature by replacing $ \rho = i t$. This analytic continuation
describes closed universe cosmologies with spatial surfaces given
by $\Sigma$ which expand from a big bang and collapse to a big
crunch, see figure 1.

\ifig\continu{In (a) we sketch
 a Euclidean geometry with two boundaries and
with a reflection symmetry around $\rho=0$.
 In (b) we sketch the lorentzian geometry that results from the analytic
 continuation $\rho = i t$. It describes a big bang to big crunch cosmology.
}{\epsfxsize 2 in\epsfbox{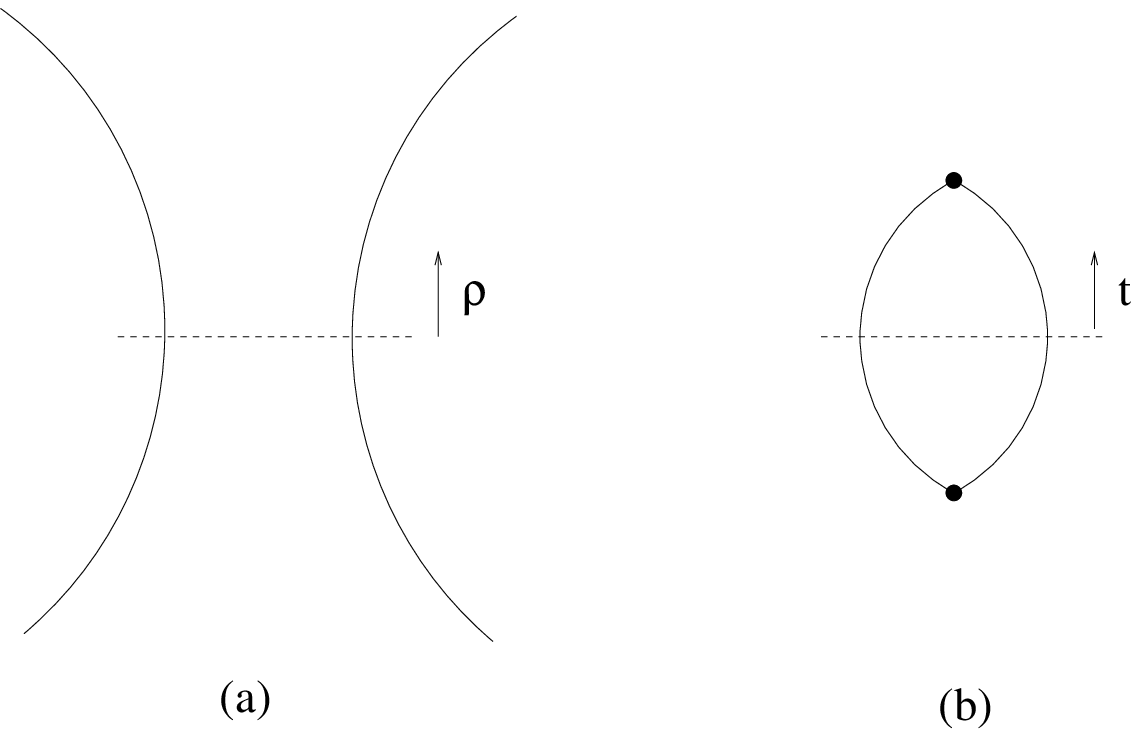}}

Our main motivation was to find the field theory interpretation
of these solutions. In particular, we wanted to find
the field theory interpretation of the closed cosmologies.
 We could not find a definitive answer for
these questions and we offer some speculations at the end of the
paper.

In most of this paper we focus on constructing various solutions
that have the general form \metric . Before we describe each
 particular case  we would like to describe various
general aspects of the solutions. Since the solutions are asymptotically $AdS$ our first goal
 will be to
understand whether the boundary conditions are stable. For example, we want to make sure that we cannot decrease
the action  by creating a brane in the interior and moving
it all the way to the boundary%\foot{In
%\ref\McInnesCosm{ B.~T.~McInnes,``A positive cosmological constant in string theory through AdS/CFT wormholes,''
%Nucl.\ Phys.\ B {\bf 609}, 325 (2001) [arXiv:hep-th/0105151].} this instability was used to claim that such
%solutions imply the existence of a braneworld with a positive cosmological constant. }
.

These are asymptotic instabilities which arise near each boundary independently and do not come
from the fact that the solution has two
boundaries. They are related to the fact that the corresponding boundary field theories are not well defined, the
boundary theory would not have an action that is bounded below. Examples of these instabilities were discussed in
\SeibWit , where an instability of this type was found when the boundary has negative curvature. In the field
theory they arise due to the conformal coupling of the scalar fields - negative curvature translates into an
effective negative mass for the scalar fields. One can go around this problem in two ways, one is to consider
positive curvature boundaries. The second is to consider a two dimensional field theory which can be defined on an
arbitrary Riemann surface.

Once we consider the two boundary solution it can happen that the
bulk Euclidean action has some negative modes. This would mean that the
solution is a saddle point but not a local minimum. Of course, the
Euclidean gravity action is not positive definite due to the
conformal factor of the metric. This is a problem even in usual
euclidean AdS space. Here we will use the  prescription in
\GHP , which consists in analytically continuing the
integral over the conformal factor to imaginary values. If the
solution is asymptotically stable, then any possible negative mode
will be localized in the interior, in the region connecting the
two boundaries. Some of the solutions we found have negative modes
and some do not.

In most of the examples we found, it is possible to find other
configurations which have the same asymptotic boundary conditions
but  consist of two disconnected spaces attached to each of
the two boundaries (see figure 2c).
 The simplest such example is to add an end of
the world brane at $\rho =0$. This end of the world brane could
arise from an orbifold or orientifold involving the $\rho$
direction and any number of internal dimensions. But the
disconnected geometries could also  be a bit more complicated. We
will present some examples below.

 Beyond perturbative
instabilities, we could ask if there is another configuration with
lower action.
In some cases  we find that there is.
Then the wormhole solution
is, at best, local minimum but not global minimum.

One can wonder if there is a supersymmetric solution
with two boundaries. We did not find any simple example. There is
a simple argument that rules out solutions of the form \metric ,
with a reflection symmetry at $\rho =0$ which analytically
continues to a time reflection symmetry. These solutions cannot be
supersymmetric since Lorentzian supersymmetric solutions must have
a timelike or null killing vector \susykill . But it is clear that
the analytic continuation of \metric\ cannot have such killing
vectors if the spatial sections are compact\foot{ If the spatial
sections are non-compact then one can have supersymmetric
solutions of this form. An example is $AdS$ space written in the hyperbolic
slicing.}.

Finally we should point out a  difference between the wormholes
we find and the axionic wormholes discussed in \sgas .
An axionic wormhole in AdS is such that the two wormhole ends can
be moved around AdS by the AdS isometries. In contrast, in
our solutions  some of
 the AdS isometries are broken by the boundary conditions in such a way
 that the wormhole ends are
localized in the central  region of the geometry, which would
correspond to the IR of the boundary theory.

\newsec{ Quotients of hyperbolic space}

The simplest example is constructed as follows
\ref\wittenprivate{
E. Witten, private communication.}.
Let us start with the Euclidean AdS (or hyperbolic space) metric
written in terms of hyperbolic slices
\eqn\hyperbol{
ds^2_{H_{d+1}}=d\rho^2+\cosh^2\rho ds^2_{H_d}
}
where $\rho\in (-\infty,\infty)$.

Even though the coordinates in \hyperbol\ seems suggestive of two
boundaries at $\rho=\pm\infty$, these spaces have only one
boundary, which is a sphere $S^d$. This can be checked by a
conformal transformation or by following the coordinate
transformation between \hyperbol\  and the sphere slicing. More
explicitly, each boundary would be a hyperbolic disk $H_d$.
However the two disks are joined at their  boundaries. In fact,
 we can
think of each disk as half of the sphere $S^d$. In other words,
when we consider a quantum field theory in hyperbolic space $H_d$,
we need some boundary conditions at the boundary of $H_d$. In this
case the boundary conditions are the following. We take two disks
and put ``transparent" boundary conditions at the boundary, i.e.
the boundary conditions we have when we consider the sphere and we
separate the sphere into two hemispheres. This situation was
studied in \ref\hyperads{ A.~Karch and L.~Randall, ``Open and
closed string interpretation of SUSY CFT's on branes with
 boundaries,''
JHEP {\bf 0106}, 063 (2001)
[arXiv:hep-th/0105132].
%%CITATION = HEP-TH 0105132;%%
M.~Porrati, ``Higgs phenomenon for the graviton in AdS space,'' Mod.\ Phys.\ Lett.\ A {\bf 18}, 1793 (2003)
[arXiv:hep-th/0306253].
%%CITATION = HEP-TH 0306253;%%
M.~Porrati, ``Higgs phenomenon for 4-D gravity in anti de Sitter space,'' JHEP {\bf 0204}, 058 (2002)
[arXiv:hep-th/0112166].
%%CITATION = HEP-TH 0112166;%%
}. Of course, we could consider the field theory in hyperbolic
space $H_d$ with other boundary conditions (i.e. non-transparent
boundary conditions). We expect that this will introduce an end of
the world brane at $\rho =0$. The precise form of this end of the
world brane will depend on the boundary conditions. For example,
we can consider the M2 brane field theory and put the boundary
conditions that result when an M2 brane is ending on the M-theory
end of the world brane. The supergravity solution is \hyperbol\
with an end of the world brane at $\rho =0$ so that we have only
one boundary. In general, this end of the world brane can be an
orbifold or an orientifold which reflects the $\rho$ coordinate
and any number of  internal coordinates. This orbifold or
orientifold should  which should be consistent with the RR fluxes
present in the system.

We can now make a
quotient of  hyperbolic slices $H_d$ in \hyperbol\ by a
discrete subgroup of the hyperbolic symmetry group, $SO(1,d)$.
We can pick this
group $\Gamma$ so that $\Sigma_d = H_d/\Gamma$ is a compact,
smooth, finite volume surface.
Now the resulting space has a metric of the form \metric\ with two
disconnected boundaries\foot{ The boundaries are connected through the
bulk, but they are disconnected through the boundary.}.
In the case $d=2$, $\Sigma_2$ is a constant
curvature Riemann surface of genus $g \geq 2 $.

The Lorentzian continuation of \hyperbol\ gives a big bang/big crunch
cosmology considered in \ref\HorowitzXK{
G.~T.~Horowitz and D.~Marolf,
``A new approach to string cosmology,''
JHEP {\bf 9807}, 014 (1998)
[arXiv:hep-th/9805207].
%%CITATION = HEP-TH 9805207;%%
}\foot{ \HorowitzXK\ proposed a description of these cosmologies
in terms of a quotient of the Lorentzian boundary theory. It is not
clear how to construct these theories. Here we are suggesting that
the cosmological solutions are somehow related to the well defined
boundary Euclidean theories.
However,   we did not specify the precise
relation.}

\subsec{ Perturbative analysis}

Here we consider potential perturbative instabilities.
We consider a metric of the form
 \eqn\coord{ ds^2 =
d\rho^2 + \cosh^2 \rho ds^2_{\Sigma_d} }
where $\Sigma_d = H_d/\Gamma$ is a constant negative curvature
compact manifold.
Let us consider scalar fields.
We will be interested in eigenfunctions  $(-\nabla^2+m^2)
\phi= \lambda \phi$ with $\lambda \leq 0$ which are normalizable.
The lowest values of $\lambda$ will be achieved for functions
that are independent of the coordinates on $\Sigma_d$.
After the  change of
variables $(1-w) = \cosh^2\rho$ the
eigenvalue equation becomes a  hypergeometric equation with
parameters
\eqn\param{ \alpha ={d\over 4} + \sqrt{ ({d\over 4})^2
+ {\tilde m^2 \over 4} }
 ~,~~~  \beta= {d\over 4} -
\sqrt{ ({d\over 4})^2 + {\tilde m^2 \over 4} } ~,~~~~ \gamma =
{1 \over 2}
}
 where $\tilde m^2 = m^2 - \lambda $.
The two solutions are $F(\alpha, \beta, {1 \over 2}; w)$ and
$w^{1/2}
F( {1 \over 2 } + \alpha, {1 \over 2 } + \beta , { 3 \over 2}; w)$.
Both are regular at $\rho =0$  in terms of the $\rho$ coordinate.
Demanding that the solution is normalizable at infinity we find
the condition
\eqn\condition{
\sqrt{ ({d \over 2})^2 + \tilde m^2 } = { d \over 2 } - n
}
where $ {d \over 2} \geq n>0$ is an integer. We can rewrite this as
\eqn\rewrite{
\lambda  = ( \Delta - { d \over 2} )^2 - ( { d\over 2 } - n)^2
 ~; ~~~~~~~\Delta = { d \over 2} +
\sqrt{ ( {d \over 2})^2 + m^2 } } We see that we should worry about possible negative modes only for relevant
operators\foot{
Negative mass square fields can sometimes have more subtle boundary
conditions which lead to an
expression for the conformal weight involving the other branch of the square root in \rewrite \ref\KlebanovTB{
I.~R.~Klebanov and E.~Witten, ``AdS/CFT correspondence and symmetry breaking,'' Nucl.\ Phys.\ B {\bf 556}, 89
(1999) [arXiv:hep-th/9905104].
%%CITATION = HEP-TH 9905104;%%
}. In these cases the first expression in \rewrite\ is
still valid (but there is a minus sign in the second).}

The reader might be puzzled by the following.
 We know that in AdS we do
not have any negative modes, then: why do negative modes arise when we write AdS in hyperbolic coordinates?. The
point  is that these negative modes are present only after we quotient the constant $\rho$ slices. If these slices
were non-compact hyperbolic spaces, as in \hyperbol\ then there would be no negative modes since the wavefunctions
need to be normalizable on these slices and this forces them to vary along these constant $\rho$ slices
\foot{There have been a few results in the mathematical literature, concerning the spectrum of the scalar
Laplacian in complete hyperbolic manifolds, in their quotients and in more general asymptotically hyperbolic
Einstein manifolds. It was shown that for the complete hyperbolic manifolds, the eigenvalues are always continuous
and greater than $d^2/4$. For quotients of such spaces, and for other asymptotically hyperbolic manifolds, there
can be also discrete eigenvalues below $d^2/4$. The exact statements and conditions for their existence can be
found in \mathHyp. }.

Let us consider some examples. Let us set $d=2$ so that
we are considering $AdS_3/CFT_2$. In particular,
we  consider the example of type IIB string theory on
$AdS_3 \times S^3 \times K3$. This theory has
fields that correspond to operators with conformal
weights $\Delta = L_0 + {\bar L}_0 = 1,2, \cdots $.
 In this case, there are
 no negative modes but
 a zero mode appears for an operator of $\Delta =1$.
This operator has spins $(1/2, 1/2)$ under $SU(2)_L\times SU(2)_R$
rotations of $S^3$. We can project it out if we quotient the
theory by a $Z_N \subset U(1)_L \subset SU(2)_L$. This quotient
does not lead to fixed points and removes the potentially
problematic field. The resulting conformal field theory at the
boundary has (4,0) supersymmetry and was studied in \ref\LarsenDH{ F.~Larsen and E.~J.~Martinec, ``Currents and
moduli in the (4,0) theory,'' JHEP {\bf 9911}, 002 (1999) [arXiv:hep-th/9909088].
%%CITATION = HEP-TH 9909088;%%
}.

As another example,  set  $d=4$, so that we have $AdS_5/CFT_4$, then $ 2
\leq \Delta  < 3$ leads to negative modes and $\Delta =3$ leads to
a zero mode. In the case of $AdS_5 \times S_5$ we have operators
with $\Delta =2,3$ in the spectrum. In this case we can analyze in
a bit more detail the action of one of the zero modes associated
to operators with $\Delta =3$. We choose the operator that gives
an equal mass to three of the four Yang Mills fermions. Then the
classical action for the corresponding field has the form \Porrati\
\eqn\classac{
L \sim { 1 \over 2} ( \partial \phi )^2 - { 3\over 8}
\left[ 3 + ( \cosh {2 \phi \over \sqrt{3}})^2 + 4
\cosh { 2 \phi \over \sqrt{3} }
\right]
}
This Lagrangian shows that the quartic order term in $\phi$ is negative,
so that what started out at as a zero mode in the classical approximation
ends up as a negative mode.

In this example one could imagine projecting out all operators
with $\Delta < 3$ by performing a quotient on the fivesphere,
$AdS_5\times (S^5/\Gamma)$. If $\Gamma$ contains elements that
have fixed points on $S^5$ then we have to worry about the extra
fields that can live at these fixed points. On the other hand, it
is possible to see that there is no subgroup of SO(6) that does
not give fixed points and, at the same time, projects out all the
$\Delta =2$ operators, which are in the ${\bf 20}$ of SO(6) \foot{
This can be shown easily for abelian subgroups. For non-abelian
subgroups the classification in \ref\mathpa{ W. M. Fairbairn, T.
Fulton and W. Klink, `` Finite and disconnected subgroups of
$SU_3$ and their application to the elementary particle physics
spectrum'', J. Math. Physics, Vol 5, 1038, (1964).} and
\ref\HananySP{ A.~Hanany and Y.~H.~He, ``A monograph on the
classification of the discrete subgroups of SU(4),'' JHEP {\bf
0102}, 027 (2001) [arXiv:hep-th/9905212].
%%CITATION = HEP-TH 9905212;%%
} shows that all non-abelian subgroups lead to fixed points. }.

Finally note that a quotient of
 Euclidean version of the solution \ref\BakJK{ D.~Bak, M.~Gutperle and S.~Hirano, ``A
dilatonic deformation of AdS(5) and its field theory dual,'' JHEP {\bf 0305}, 072 (2003) [arXiv:hep-th/0304129].
%%CITATION = HEP-TH 0304129;%%
} gives us a configuration where the value of the dilaton in the two
asymptotic boundaries is different.

\subsec{ Non perturbative instabilities}

For these manifolds further instabilities arise  due to possible
brane creation. These are backgrounds with a $p$-form field
strength, so we should worry about the possibility of creating
branes that screen, or partially screen, this fieldstrength. This
problem was analyzed in \SeibWit . We now summarize that
discussion. We consider a Euclidean  $(d-1)$-brane wrapping the
$d$ dimensional slices of the metric \hyperbol . We assume that
the brane is charged under the same $(d+1)$-form potential whose
field strength supports the background. In units where the AdS
radius is one its action can be written as \eqn\actionform{ S \sim
({\rm Area} - d \, { \rm Volume}) \sim (\cosh \rho)^d - d
\int_0^\rho d\rho' (\cosh \rho')^d \sim  - { 2 d \over 2^d (d-2) }
e^{(d-2) \rho} + \cdots } where we extracted the leading
dependence for large $\rho$. We see that for large $\rho$ the
leading contribution to \actionform\ is negative, so that if we
create a brane antibrane pair and we move one of them to $\rho \to
\pm \infty$ we can decrease the action. In fact, for large $\rho$,
the position of the brane can be viewed as a conformally coupled
scalar from the field theory point of view. Since the
field theory is on a negatively curved space this conformal
coupling leads to an effective negative mass term for the scalar,
which causes an instability. This instability under brane creation
is a non-perturbative instability, in the sense that we need to
overcome a barrier of order ${1 \over N} \sim { 1 \over g_s}$ in
order to create the branes. This discussion holds for situations
where we have compactified the constant $\rho$ sections. If we do
not compactify them then the negative mass of these scalar fields
is not a problem since a conformally coupled field obeys
 the Breitenlohner-Freedman
bound in $d$ dimensions. In fact, for a conformally coupled scalar
in hyperbolic space, $H^d$,  we have $R^2 m_{conf}^2 - R^2
m^2_{BF} ={1 \over 4}$ where $R$ is the radius of hyperbolic
space, $m_{conf}$ is the effective mass that results from the
coupling to the constant negative curvature for a conformally
coupled scalar and $m_{BF}$ is the Breitenlohner-Freedman bound.

Note that the instability that we are discussing here has nothing to do with the fact that the manifold has two
boundaries, it is purely due to the asymptotic geometry near each of the boundaries. It would be present as long
as the boundary has negative curvature and the spacetime solves Einstein's equations \SeibWit . So in these cases
the CFT is not well defined and we cannot find a minimal action spacetime configuration with these boundary
conditions. See \ref\McInnesXP{ B.~T.~McInnes, ``Topologically induced instability in string theory,'' JHEP {\bf
0103}, 031 (2001) [arXiv:hep-th/0101136].
%%CITATION = HEP-TH 0101136;%%
} for a more general discussion on the boundary manifolds which lead
to instabilities of this type.

One would expect that adding a large enough mass term to the
Lagrangian would remove the instabilities. For example, in $AdS_5
\times S^5$ we can turn on the operator that turns on a mass for
three of the fermions. This also induces a mass for the scalars by
supersymmetry. When the field theory is in flat space these
deformations were considered in \refs{\Porrati,\jpms}. One could
repeat their analysis on a negatively curved hyperbolic space. It
could be that the fivebranes that appear in flat space \jpms\ will
also appear here and disconnect the two asymptotic regions.

\subsec{ Correlation functions}

We can compute the  correlation functions of boundary operators
using gravity. Before doing the quotient we can compute the bulk
to boundary propagator in hyperbolic  coordinates and we find
\eqn\bulkbound{ G(r, \rho ; r_0) \sim { 1 \over (\cosh \rho)^\Delta
\left[ \cosh s -  \tanh \rho \right]^\Delta  } } where $(r, \rho)$
  label a point in the bulk, $r_0$ labels a point in the
boundary and $s$ is the distance between the points $r$ and $r_0$
 measured with
the boundary hyperbolic metric.

Now we can compute the two point function by taking $\rho \to
\infty$, renormalizing by a factor of $e^{ \Delta \rho}$ to obtain\foot{
The correct method is a bit more involved, but
this gives us the correct result up to a $\Delta$ dependent factor
  \Freedman , \ref\KostasRen{S.~de Haro, S.~N.~Solodukhin and K.~Skenderis,
``Holographic reconstruction of spacetime and renormalization in the  AdS/CFT correspondence,'' Commun.\ Math.\
Phys.\  {\bf 217}, 595 (2001) [arXiv:hep-th/0002230].\ M.~Bianchi, D.~Z.~Freedman and K.~Skenderis, ``Holographic
renormalization,'' Nucl.\ Phys.\ B {\bf 631}, 159 (2002) [arXiv:hep-th/0112119]..\ K.~Skenderis, ``Lecture notes
on holographic renormalization,'' Class.\ Quant.\ Grav.\  {\bf 19}, 5849 (2002) [arXiv:hep-th/0209067].}.}
  \eqn\twosame{ \langle O(r,
\theta^i)_1  O(r', {\theta'}^i)_1 \rangle \sim
 { 1 \over \left[\sinh s/2
\right]^{2 \Delta }} } for operators on the same boundary and
\eqn\twoother{ \langle O(r, \theta^i)_1  O(r', {\theta'}^i)_2
\rangle \sim { 1 \over \left[
 \cosh s/2 \right]^{2 \Delta }} } for operators
on opposite boundaries, where $s$ is the distance between the two
points measured with the boundary hyperbolic metric. This is the
result for non-compact hyperbolic slices. If we want to consider
the theory on the quotient $H_d/\Gamma$ we need to sum over all
images\foot{The sum over images gives us the result when we
neglect interactions in the interior. Once we take into account
interactions the result is not given by the sum over images.
Correspondingly,
 the full gauge theory correlators are not given
by a sum over images since the field theory is an interacting field theory.}. This sum over images might diverge.
By estimating the sum as an integral one can see that the sum  converges if $\Delta > d-1$. It is interesting that
this is the same condition that eliminates zero and negative modes for the field in the interior (see \rewrite ).
We see that after summing over images the  correlator across the two boundaries \twosame\ will generically give a
non-zero answer with some coordinate dependence.

Note that if the operators in \twosame\ carry any charge under a
global symmetry then these correlators vanish due to the fact that
there is a gauge field in the bulk whose Gauss's law prevents any
charge transfer across boundaries. In other words, even though the
correlators computed as in \twoother\ are nonvanishing the full
result vanishes once we integrate over the gauge field in the bulk
that is associated to the global symmetry in the boundary theory.
The net result is that there is no charge transfer among the two
boundaries.

\subsec{$AdS_3$ and the change in moduli of Riemann surfaces}

In this subsection we consider the three dimensional case (though
similar remarks might apply to higher dimensional cases). If we
have a 2d Riemann surface of constant curvature  this surface will
have moduli $t^\alpha $ which specify its shape. The moduli of the
2d Riemann surface can be different on the two sides. In fact
there is a theorem (``the Bers simultaneous uniformization
theorem"), which states that the solutions are in one to one
correspondence with a pair of points in Teichmuller space \Bers.
These two points are the values of the moduli of the 2d Riemann
surface at the two boundaries. We can represent a Riemann surface
as the  quotient of $H_2$ by a Fuchsian discrete group $\Gamma
\subset SL(2,R)$.   If we quotient $H_3$ by the same group,
$\Gamma$, now viewed as a subgroup of $SL(2,C)$ we end up with a
geometry where the moduli of the Riemann surface are the same on
the two boundaries. Quotienting $H_3$ by so called
``quasi-Fuchsian" groups gives us three dimensional spaces that
join Riemann surfaces with different moduli.

In addition, given a Riemann surface on the boundary, it is possible to find a geometry that ends on it and has no
other boundary. This is a geometry that results by quotienting $H_3$ by a so called Schottky group, see
\ref\KrasnovZQ{ K.~Krasnov, ``Holography and Riemann surfaces,'' Adv.\ Theor.\ Math.\ Phys.\ {\bf 4}, 929 (2000)
[arXiv:hep-th/0005106].
%%CITATION = HEP-TH 0005106;%%
} for further discussion\foot{See also \ref\Thurston{ W.~P.~Thurston, ``The Geometry and Topology of Three-Manifolds'',
 lecture notes, Princeton University, 1980, http://www.msri.org/publications/books/gt3m/ }
 for more about 3-manifolds and their different quotients.}.
It would be nice to see if the geometry with disconnected boundaries has larger or smaller action than the
geometry which connects the two boundaries \foot{Some Lorentzian versions of these different quotients were
described in \ref\LorBTZ{ S.~Aminneborg, I.~Bengtsson, D.~Brill, S.~Holst and P.~Peldan, ``Black holes and
wormholes in 2+1 dimensions,'' Class.\ Quant.\ Grav.\ {\bf 15}, 627 (1998) [arXiv:gr-qc/9707036] .\ S.~Aminneborg,
I.~Bengtsson, S.~Holst and P.~Peldan, ``Making Anti-de Sitter Black Holes,'' Class.\ Quant.\ Grav.\  {\bf 13},
2707 (1996) [arXiv:gr-qc/9604005].\ M.~Banados, ``Constant curvature black holes,'' Phys.\ Rev.\ D {\bf 57}, 1068
(1998) [arXiv:gr-qc/9703040].\ K.~Krasnov, ``Analytic continuation for asymptotically AdS 3D gravity,'' Class.\
Quant.\ Grav.\  {\bf 19}, 2399 (2002) [arXiv:gr-qc/0111049].}.}.

\ifig\twobdy{In (a) we see a three dimensional geometry with two
boundaries. Both boundaries are identical Riemann surfaces. These
results from modding out $H_3$ by a Fuchsian group. In (b) we see
a three dimensional manifold ending on two Riemann surfaces with
different values of the moduli. These result from quotienting
$H_3$ by a quasi-Fuchsian group. Finally in (c) we see two
disconnected three dimensional manifolds each ending on one
boundary. Each of these manifolds results from quotienting $H_3$
by a Schottky group.}
{\epsfxsize 2 in\epsfbox{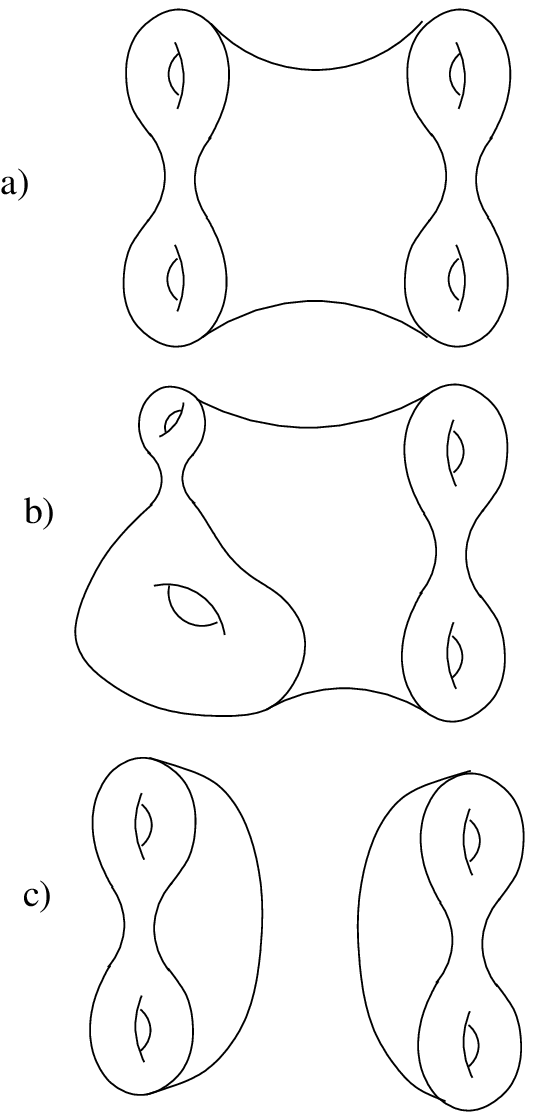}}

\subsec{ A rather stable example}

Let us  consider an $AdS_3/CFT_2$ example.
An ordinary two dimensional
conformal field theory can be defined on any Riemann surface.
On the other hand we saw above that there is generically an instability
under brane creation. In the two dimensional case the computation in
\actionform\ gives an action going like $S \sim - \rho $,  for large $\rho$.
We should note however that the computation in \actionform\ assumed the
existence of a D-brane in the bulk whose tension, $T$, is equal
to its charge, $q$, (in some units), so that the leading contribution to
\actionform\ cancels. In supersymmetric backgrounds we have  a BPS bound
ensuring that $T \geq q$. However there is no particular reason why
there should exist a brane that saturates this equality.

Let us be more concrete. Consider the example of $AdS_3 \times S^3
\times K3$. Let us  consider this system with $Q_1$ units of NS
electric flux on $AdS_3$ and $Q_5$ units of NS magnetic flux on
$S^3$. If the values of all RR fields  on the $K3$ vanish then  a
fundamental string (or an NS fivebrane) has $T=q$ and leads to an
instability as discussed above. In \SeibWit\ it was observed that
the dual conformal field theory is singular since its target space
is non-compact. This target space can be thought of as the moduli
space of $Q_1$ instantons of $SU(Q_5)$ gauge theory on $K3$. This
non-compact region comes from small instanton singularities. If we
turn on some particular RR fields on $K3$ then the instantons
become non-commutative instantons and the small instanton
singularity is removed. In fact, this is the situation at  generic
point in the moduli space of the boundary CFT if $Q_1$ and $Q_5$
are coprime.   In this situation we find that all branes in
$AdS_3$ will have $T
> q$.
If $Q_1$ and $Q_5$ are large then we can find branes whose charges
 are closely aligned with $Q_1,Q_5$.
  In this case the action
\actionform\ becomes \eqn\actionmorecl{ S = T({\rm Area}) - 2 q
({\rm Volume}) = T (\cosh \rho)^2 - { q \over 2} \sinh 2 \rho - q
\rho  \sim { q \over 4} [ \epsilon  e^{ 2 \rho} - 4   \rho + 2] }
with a  small $\epsilon = { T - q \over q}
>0$. We see that for large $\rho$ the action \actionmorecl\ is
positive. This means that the asymptotic boundary conditions are
stable. We can mod out the theory by a $ Z_N \subset U(1)_L
\subset SU(2)_L \subset SO(4)$, where SO(4) is the group of
rotations of $S^3$. Then it is possible to remove all perturbative
instabilities. The dual CFT corresponds to the theory we obtain if
we consider D1 and D5 branes at an $A_N$ singularity, which was
studied in \LarsenDH .
So in this case the two boundary solution is a perturbatively stable solution.

On the other hand, it is clear that for large $Q_1$, $Q_5$ it will be possible to find branes with small
$\epsilon$. If $\epsilon \ll 1$ then we see that the action \actionmorecl\ becomes negative for some intermediate
values of $\rho$, even though it is positive for large $\rho$. So we get a non-perturbative instability under
brane creation, i.e. we can decrease the action of the Euclidean solution by creating brane/anti-brane pairs and
moving them to a suitable position in the $\rho$ coordinate where \actionmorecl\ is negative. Notice that we can
only reduce the action by a finite amount through this process. So we expect to find another solution which will
indeed be stable.

Of course this system also has the solutions depicted in figure 2 (c). But
we did not compute the action difference between the solutions in figures
2(a) and 2(c).

In summary, in this example we have a well defined conformal field theory
which can live on any Riemann surface. We find a simple geometry
that connects two Riemann surfaces. This geometry seems to be
perturbatively
stable, though it has non-perturbative instabilities under brane creation.

\newsec{An example based on merons}

In this section  we construct wormholes that connect two $AdS_4$
regions that involve $SU(2)$ gauge fields. These were considered
in \Hosoya \ref\Reymerons{ S.~J.~Rey, ``Space-Time Wormholes With Yang-Mills Fields,'' Nucl.\ Phys.\ B {\bf 336}, 146
(1990).\ S.~J.~Rey, ``The Confining Phase Of Superstrings And Axionic Strings,''
Phys.\ Rev.\ D {\bf 43}, 526 (1991).} \ref\GuptaBS{ A.~K.~Gupta, J.~Hughes, J.~Preskill and
M.~B.~Wise, ``Magnetic Wormholes And Topological
Symmetry,'' Nucl.\ Phys.\ B {\bf 333}, 195 (1990).
%%CITATION = NUPHA,B333,195;%%
} \foot{ In \GuptaBS\ it was argued that these meron configurations do
not contribute in four flat dimensions. As we explain later, they do not
contribute in $AdS$ if we have the standard boundary conditions for the
gauge fields. On the other hand they do contribute if we impose special,
non-standard boundary conditions. We will consider this latter case below.}.
Here we will embed these solutions in M-theory and then we will discuss
some aspects of the solution.

We start with eleven dimensional supergravity compactified on $S^7$. This has a
 well known solution of the form
$AdS_4 \times S^7$ whose dual field theory is the theory on coincident $M2$ branes or IR limit of 2+1 Yang Mills
with sixteen supercharges. The supergravity theory has a consistent truncation to an SO(8) gauged supergravity
theory \ref\deWitIY{ B.~de Wit and H.~Nicolai, ``The Consistency Of The S**7 Truncation In D = 11 Supergravity,''
Nucl.\ Phys.\ B {\bf 281}, 211 (1987).
%%CITATION = NUPHA,B281,211;%%
}. This in turn has a truncation to a theory with just an SU(2)
gauge field and a graviton. This truncation can be understood as
follows. We consider the spinor representation ${\bf 8}_c$ of
SO(8). SO(8) triality maps this to the vector ${\bf 8}_v$ of
SO(8). We consider the $SO(3) \times SO(5)$ subgroup that rotates
the first three and the second five directions of this vector. The
truncation consists in keeping all fields invariant under this
SO(5). This projects out all the scalars of $SO(8)$ gauged
supergravity \ref\RomansQI{ L.~J.~Romans, ``The SO(5) Sectors Of
Gauged N=8 Supergravity,'' Phys.\ Lett.\ B {\bf 131}, 83 (1983).
%%CITATION = PHLTA,B131,83;%%
}. The bosonic fields are just the graviton and
the Yang Mills fields. Another way to obtain this truncation is first
to consider the SO(4) truncation studied in
\ref\Cveticfour{M.~Cvetic, H.~Lu and C.~N.~Pope,
``Four-dimensional N = 4, SO(4) gauged supergravity from D = 11,''
Nucl.\ Phys.\ B {\bf 574}, 761 (2000)
[arXiv:hep-th/9910252].
%%CITATION = HEP-TH 9910252;%%
} and then further truncate it to the diagonal $SU(2)$ in
$SU(2)_L \times SU(2)_R= SO(4)$.  It is useful to understand how this $SU(2)$
subgroup acts on the original vector representation of $SO(8)$, i.e.
the representation which the scalars on the M2 brane belong to. If
we divide the 8 scalars into 4+4  we break SO(8) to $SO(4) \times SO(4)$.
Each of the SO(4) factors is a product of two SU(2) groups. The SU(2)
we are interested in is the diagonal combination of an SU(2) coming from
each of the two SO(4). In other words the ${\bf 8 }_v$ of
SO(8) transforms a pair of ${\bf 2}$s of SU(2).

In summary, we end up with the action
\eqn\gravfourd{
S \sim N^{3/2} \int d^4x \sqrt{g}[-(R+6)+ \alpha F^a_{\mu\nu}F^{a\mu\nu}]
}
with $\alpha =1$.
We have chosen units such that the radius of $AdS$ is one for the
solution with $F=0$, and the gauge field is normalized so that
the connection is given by $A = i {\sigma^a \over 2} A^a $ and $F = dA + A^2$.

\subsec{A simple one boundary  solution}

If we take a gauge field such that $F=\pm *F$ , then its stress
 tensor is zero and the geometry is still $AdS_4$.
One way of viewing this solution is the following. $AdS_4$ is
conformally flat. Since the self duality  equations are
independent of the conformal factor we can take a solution in flat
space and translate it to a solution in $AdS_4$. There is,
however, an important point. $AdS_4$ is conformal to the interior
of a unit ball in $R^4$. This implies that with the generic
instanton configuration the gauge field at the boundary of the
unit ball will not vanish, more precisely, it will not be pure
gauge. This means that the gauge field does not vanish at the
boundary of $AdS$. Therefore the dual field theory is not the
usual field theory on coincident M2 branes. It is the   M2 brane
field theory which has been coupled to a fixed background SU(2)
gauge connection. This connection couples to the SU(2) currents
which are part of the SO(8) global symmetry currents of the
theory. In appendix A.6 we write the action for a single M2
brane\foot{ There is no Lagrangian for multiple M2 branes. It is
not clear how to write a lagrangian for 2+1 super YM that will
flow to the one we are considering since the SU(2) subgroup we are
dealing with is not contained in SO(7), which is the symmetry
group of $2+1$ SYM with 16 supercharges.}. Even though an
instanton preserves supersymmetry in flat space, once we are in
AdS it no longer preserves supersymmetry. Furthermore, the boundary
conditions themselves break supersymmetry. So we are dealing with
a non-supersymmetric deformation of the coincident M2 branes
theory. In appendix A.5 we prove these assertions.

If we consider a single instanton at the center of $R^4$,
 this gives us  a configuration that is spherically
symmetric, up to  gauge transformations.
The  gauge field can be written as
\eqn\bcgauf{ A^a = f w^a
}
where $w^a$ are the left invariant one forms on $S^3$ and $f$ is a function of the radial coordinate, see
appendix A.1 for more details. The self duality conditions for the instanton imply that in $R^4$ $f$ obeys a
first order equation that is solved by setting $f^{-1} = 1 + {r_0^2 \over r^2}  $, where $r$ is the radial
coordinate on $R^4$. In terms of the $AdS_4$ metric $ds^2 = d\rho^2 + \sinh^2 \rho ds_{S^3}^2 $ we find that
\eqn\finf{ f(\rho) = { f_B \sinh^2\rho/2 \over \cosh^2\rho/2 - f_B} } When $\rho \to \infty$ this asymptotes to
the boundary value $f_B $. When $f_B = 1/2$ we have precisely half an instanton in $AdS$.

For each value of the boundary condition, $f_B$, there are at least two possible gauge field configurations in the
interior. One corresponding to a instanton and one corresponding to an anti-instanton in $R^4$ with the same
boundary value of the fields at the unit sphere in $R^4$. The action for these self dual configurations can be
computed in terms of the instanton charge inside $AdS$ and it is equal to \eqn\actinst{ S_{sd}[f_B] = \alpha {\cal
N} 16 f_B^2(3 - 2 f_B) } for the self dual configuration, where ${\cal N}$ is an overall factor common to all
computations we are doing\foot{ ${\cal N}$ is equal to the normalization factor of \gravfourd\ outside the
integral times the volume of $S^3$, $\Omega_3 = 2 \pi^2$.}. For the anti-self dual configuration we find
$S_{asd}[f_B] = S_{sd}[1-f_B]$. In the particular case $f_B=1/2$ both are equal. The gravity part of the action is
equal to $S_{grav} =4 {\cal N}$ after subtracting the usual counter terms, see appendix A.2.
 So the total action for the $f_B=1/2$ case is
\eqn\actoneb{
S_{1~bdy} = ( 4 + 8 \alpha) {\cal N}
}

An interesting configuration is the zero size instanton. In this
case it seems that the solution might be supersymmetric (we are
not sure because we cannot trust supergravity). It is also a
solution with the same boundary conditions as the usual $AdS_4
\times S^7$ solution. It would be very nice to figure out the role
that these zero size instantons play in the physics of $AdS_4
\times S^7$. If we start with $AdS_4 \times S^7$ with $f_B=0$ and
we increase continuously $f_B$ to $f_B =1$, then we end up with a
zero size instanton in the interior. On the other hand $f_B=1$ is
gauge equivalent to $f_B=0$. So  this sum over zero size
instantons is necessary to make sure that the field theory is
invariant under large gauge transformations.

\subsec{A two-boundary  solution}

Now let us consider the situation where the boundary values of the
gauge fields are given by \bcgauf\ with $f_B=1/2$. It turns out
that in this case we can look for a solution with $f(\rho) = 1/2$.
This will solve the Maxwell equations, as long as the metric is
SO(4) symmetric. This is not a self dual configuration, so the
stress tensor is non-zero and one should solve Einstein's
equations. We find that the solution is given by \eqn\soleinst{
ds^2 = d\rho^2 + e^{2 \omega} ds_{S^3}^2 ~,~~~~~~~~~~ e^{2 \omega}
= \sqrt{ \alpha + { 1 \over 4} } \cosh 2 \rho - {1 \over 2} } In
this solution the three sphere never shrinks completely, it has a
minimum size at $\rho =0$ and it increases as $\rho \to \pm
\infty$.

A natural question is whether this two boundary solution is stable
under small perturbations. A good candidate for an unstable mode
is a mode of the form \eqn\deformat{ A=\left[{1 \over 2} +\epsilon
(\rho)\right]gdg^{-1}\;\; ~~~~ {\rm with } \;\;\;\;,\;\;
\lim_{t\to \pm\infty}\epsilon(\rho) = 0 } where $\epsilon$ is
small, and $g$ is an $SU(2)$ group element, such that
$i{\sigma^a\over 2}w^a=g^{-1}dg$. Inserting this into the
Euclidean action and expanding to second order in $\epsilon$ we
find that indeed there is a negative mode. See appendix A.3  for
details.

It is interesting to find the action for this configuration. For $\alpha=1$
it
approximately given by $S = (30.296...){\cal N}$.
This action is larger than   twice the action of the one boundary solution
\actoneb .
This suggests that the instability we saw above will deform the two
boundary
solution into two disconnected two boundary solutions.

It is interesting to note that for other values of $\alpha$ in the action
\gravfourd\ the physics is different. If $\alpha > 1.245 \pm .005$ then
the two boundary solution is perturbatively stable. Furthermore the
action of the two boundary solution will be smaller than the action of
two one boundary solutions when $ \alpha > 3.775 \pm .005 $.
We do not know of a configuration in string theory that would give
\gravfourd\ with these large values of $\alpha$.

\newsec{An example involving $N=4$ Yang Mills}

In this section we will discuss an example that is physically rather
similar to the one we found above. This example will produce again
a two boundary solution, however in this case we do not find any negative modes around the solution, so it might be a local minimum of the Euclidean
functional integral. The basic idea is to start with the standard
$AdS_5 \times S^5$ example of AdS/CFT. Suppose we could
 truncate the supergravity
theory to a 5 dimensional theory involving an SU(2) gauge field and the metric.
 We could consider a
configuration containing an instanton on the $S^4$ boundary of
$AdS$. This would   lead to a two boundary solution similar to the
one we had above. It turns out that it is not possible to perform
this truncation due to the presence of Chern Simons terms in the
five dimensional theory. Instead, we will consider the consistent
KK reduction to five dimensions considered in \ref\Cveticfive{
M.~Cvetic, H.~Lu, C.~N.~Pope, A.~Sadrzadeh and T.~A.~Tran,
``Consistent SO(6) reduction of type IIB supergravity on S(5),''
Nucl.\ Phys.\ B {\bf 586}, 275 (2000) [arXiv:hep-th/0003103].
%%CITATION = HEP-TH 0003103;%%
}. This is a reduction that keeps the SO(6) gauge fields and some of the scalar fields parameterizing deformations
of the five-sphere. The  reduced action is \eqn\redact{\eqalign{ {\cal L}_5 &= R\, *1 - {1\over 4}
T^{-1}_{IJ}\, *D T_{JK}\wedge T^{-1}_{KL}\, DT_{L I} - {1\over 4}T^{-1}_{IK}\, T^{-1}_{JL}\, * F^{IJ}\wedge F^{KL}
-V\, *1 \cr & \!\!\! - {1\over 48} \epsilon_{I_1\cdots I_6} \Big(F^{I_1 I_2} F^{I_3 I_4} A^{I_5 I_6} -
 g F^{I_1 I_2} A^{I_3 I_4} A^{I_5 J} A^{J I_6}
+{2\over 5} g^2 A^{I_1 I_2}  A^{I_3 J} A^{J I_4} A^{I_5 K} A^{K I_6} \Big) }}
%%%%%
where $A^{IJ}$ are the $SO(6)$ gauge fields, $T^{IJ}$ is a $6\times 6$ symmetric unimodular matrix of scalars and
the potential $V$ is given by $ V = {1\over 2} g^2\, \Big(2 T_{IJ}\, T_{IJ} - (T_{II})^2 \Big)\,. $

Consider now the $SO(3) \times SO(3)$ subgroup of $SO(6)$ that
rotates the first three coordinates and the second three
coordinates of $R^6$. We consider a configuration that consists of
an instanton on $S^4$ for the first $SO(3)$ and an anti-instanton
for the second $SO(3)$. We further consider instantons and
anti-instantons that are $SO(5)$ symmetric under rotations of
$S^4$. This gauge field configuration sets the boundary conditions
at the boundary of $AdS_5$. As in the example we considered above,
the corresponding gauge theory is ${\cal N} =4 $ SYM with an
external fixed gauge field coupled to the $SO(6)$ currents. In
this case the field theory itself, as defined on each boundary
seems stable and well defined.

Let us describe more concretely the two boundary solution.
We consider a five
dimensional space which is SO(5) symmetric and it is foliated by $S^4$s.
In
other words,
 the metric has the form
 \eqn\metrifi{ ds^2 = d\rho^2 + e^{ 2 \omega} ds^2_{S^4}=
 d\rho^2 + e^{ 2 \omega} ( d \theta^2 + \sin^2 \theta \,
 { 1 \over 4} w^a w^a)
} where $w^a$ are the left invariant one forms on $S^3$.
 We pick a gauge
field configuration given by
\eqn\gaguef{ A_\mu^{IJ} = i A_\mu^a L^{a \, IJ} + i \tilde A^a_\mu {\tilde L}^{a \,
IJ} }
where $L^a$ and ${\tilde L}^a$ are the generators of $SO(3) \times SO(3)$.
 We pick a gauge field $A^a$ to be an
instanton and ${\tilde A}^a$ to be an anti-instanton
which are $SO(5)$ symmetric under rotations of $S^4$
\eqn\gfields{ A^a = \cos^2
{\theta \over 2}  w^a ~,~~~~~~~ {\tilde A}^a = \sin^2 {\theta
\over 2}w^a }
Let us check that this is a configuration  which solves the equations
of motion for the scalars in the {\bf 20} of SO(6) that couple to
the Yang Mills fields.
The linearized coupling to these scalars
has the form \eqn\couplform{ \phi_{IJ} F^{IK}_{\mu \nu} F^{KJ \,
\mu \nu } } where the indices of $\phi_{IJ}$ are symmetric and
traceless. On the other hand, the configuration under
consideration was designed so that $F^{IK}_{\mu \nu} F^{KJ \, \mu
\nu } \sim \delta^{IJ}$. Then \couplform\ vanishes.

The equation of motion for the Yang Mills fields involve a term coming from the $F^2$ term in the Lagrangian and
one term coming from the Chern-Simons coupling in \redact\ : $\epsilon_{I_1 \cdots I_6} F^{I_1 I_2} \wedge F^{I_3
I_4} \wedge A^{I_5 I_6} + \cdots $. This leads to a term in the equations of motion for $A^{IJ}$ of the form
\eqn\termcs{ { \delta S_{CS} \over \delta A^{IJ} } \sim \epsilon_{IJKLMN} F^{KL}\wedge F^{MN} } Our choice of
instanton and anti-instanton configuration has \eqn\choice{ F^{KL}\wedge F^{MN}  \sim (
 L^{a~KL}L^{ a~MN} - {\tilde L}^{a ~KL} {\tilde L}^{a~MN})\epsilon_4
}
When this is inserted in \termcs\ we get zero.  See appendix A.7 for details.
Note that \choice\ implies that the total $SO(6) $
instanton number is zero.

If we were to choose ${\tilde F}^a = F^a$ then we would get a
cross term in \choice\ that involves $L^{a \, KL} {\tilde L}^{a \,
MN}$ which would lead to a non-zero right hand side in \termcs .
This choice of gauge fields leaves an unbroken $U(1)$ that rotates
the first three coordinates in $R^6$ into the second three
coordinates of $R^6$. The nonzero result for \termcs\ is implying
that the instantons generates a charge under this U(1). It might
be possible to add other charged particles to the system which
will cancel this extra term. This seems to complicate the analysis
and we have not explored this possibility.

Now we need to find the five dimensional geometry. We can write the
effective action for $\omega$ in \metrifi\
\eqn\effact{
S\sim  \int d\rho e^{ 4 \omega} ( -{\omega'}^2 - e^{- 2 \omega} -1 + \beta e^{-
4 \omega} )
}
where in our case $\beta = 1$.
We need to consider the zero energy solution,
which turn out to be identical
to the one in the previous section. So
 the metric is given again by
\eqn\metricf{
e^{2 \omega} = \sqrt{\beta + {1 \over 4} } \cosh 2 \rho  - { 1 \over 2}
}

Note that the gauge field on $S^4$ is topologically trivial as an
SO(6) gauge field. In fact, in appendix A.8 we describe a continuous
deformation which transforms the instanton/anti-instanton configuration
into a configuration which is a gauge transformation of the zero
gauge field configuration.
So we expect that if we consider fluctuations
of the gauge fields that carry indices under both $SO(3)$ groups we will
get negative modes. In fact, if we consider the SO(6) gauge theory
on $S^4$, we can see that the instanton/anti-instanton configuration
has some negative modes. We find a negative mode explicitly in appendix A.8.
However, it turns out that when we include the radial direction and
we demand that fluctuations are normalized at $\rho = \pm \infty$, then
we find that the  negative mode on $S^4$ does
not give rise to negative modes in the full five dimensional geometry.

We also studied possible negative modes coming from the scalars in the
${\bf 20}$ of SO(6) and we found that there were none, see appendix
A.9.
 The conformal
factor of the metric does not  lead to negative modes either,
once we Wick rotate it as in \GHP .
In principle we should check all other fields of the ten dimensional
supergravity theory to check that there are no negative modes coming
from other fields.
We only  analyzed the fields that we thought were natural candidates
for negative modes.

In this case it is also likely that there is a single boundary
solution, but we did not find it explicitly.

\newsec{ Discussion}

In this paper we have considered a
 variety of Euclidean solutions which connect two boundaries. These pose a
puzzle from the point of view of AdS/CFT. From the field theory point of
view the correlation functions across the two boundaries should
factorize, while from the gravity point of view they do not.
So we can see two possibilities. Either one can introduce a
subtle correlation between the two field theories or
 the full quantum gravity
  answer, after summing over all geometries, is such that  the
gravity correlators indeed factorize.

We have emphasized the presence or absence of perturbative
negative modes. The simplest case is when there are no
perturbative negative modes. Then the two boundary
geometry represents a local minimum of the action. If a negative
mode exists, then there are nearby configurations which might
contribute in a more important way to the partition function than
the configuration we started with. In this case it is less clear
that there is indeed a correlation between the two boundaries
(thought it is also not clear that there is no correlation).

If we take two decoupled field theories then we are
instructed to sum over all manifolds that end on the respective boundaries.
Why should we  forbid the manifolds that connect the two boundaries?
If the two field theories are truly decoupled we would conclude that the
final effect of performing the sum over all manifolds that connect the
two boundaries would completely factorize into the two partition functions
of the two field theories.

In the context of ordinary wormholes the corresponding problem is the fact that the wormholes seem to induce
non-local interactions
that would lead to violations of unitarity in the Lorentzian theory
\ref\stromch{ A. Strominger, `` Vacuum Topology and Incoherence in Quantum
Gravity", Phys. Rev. Lett. {\bf 52}, 1733 (1984).}\ref\Qcoh{
G.~V.~Lavrelashvili, V.~A.~Rubakov and P.~G.~Tinyakov, ``Disruption Of Quantum Coherence Upon
A Change In Spatial Topology In Quantum Gravity,'' JETP Lett.\  {\bf 46}, 167 (1987) [Pisma Zh.\ Eksp.\ Teor.\
Fiz.\  {\bf 46}, 134 (1987)].\ G.~V.~Lavrelashvili, V.~A.~Rubakov and P.~G.~Tinyakov, ``Particle Creation And
Destruction Of Quantum Coherence By Topological Change,'' Nucl.\ Phys.\ B {\bf 299},
757 (1988).}\ref\hawk{ S. W. Hawking, ``Quantum Coherence down the Wormhole",
Phys. Lett. {\bf B195} 337 (1987).}.
By an
``ordinary wormhole" we mean a wormhole such as the axionic wormhole of \sgas\ which could connect any two points
in spacetime. In this context a possible solution was suggested by Coleman \ref\ColemanCY{ S.~R.~Coleman, ``Black
Holes As Red Herrings: Topological Fluctuations And The Loss Of Quantum Coherence,'' Nucl.\ Phys.\ B {\bf 307},
867 (1988).
%%CITATION = NUPHA,B307,867;%%
}. He suggested that
the effect of the sum over all possible wormholes leads to
superselection sectors characterized by some parameters $\alpha_i$.
These  parameters characterize the wavefunction of the closed
universes associated to wormholes. An observer would see local
and unitary physics but she would not be able  to compute
 the parameters $\alpha_i$ from first principles, she would have
 to measure them experimentally.
 An  axionic wormhole then would look like an ordinary
instanton effect that breaks the U(1) translation symmetry of the axion.
In an AdS solution to string theory   where axionic wormholes existed one
would be forced to conclude that the local field theory at the
boundary of AdS corresponds to the theory with fixed $\alpha_i$
parameters, i.e. to a given superselection sector\foot{
We do not know of a concrete nonsingular string theory AdS example where axionic wormholes actually
do exist.}. Otherwise the integral over $\alpha_i$ would induce non-local
effects in the field theory.

It is tempting to speculate that one could have  a similar situation with the wormholes that we found in our
paper. These wormholes  would have a large number of associated $\alpha$ parameters. The number of parameters is
the dimension of the Hilbert space for the associated compact universe. This can be estimated as a typical area in
the compact universe, which in our cases goes as the central charge of the corresponding field theory\foot{ In the
supergravity approximation all definitions of the central charge give similar values. In the examples of sections
3,4,5 this goes as $c \sim Q_1 Q_5 , ~ N^{3/2}, ~ N^2$ respectively. The number of states then would be of the
order of $e^c$. }. This would give a large number of parameters for large $N$. In Coleman's arguments it was
important that the wormholes could exist with  arbitrary wavefunctions and this leads to an indeterminacy of the
$\alpha$ parameters \ColemanAlpha . If one thinks that the field theories that we introduced correspond to
particular values of $\alpha$ parameters, then it is hard to see which deformations of the field theories would
lead to other values\foot{ See
\ref\Reyalpha{S.~J.~Rey, ``Holographic principle and topology change in string
theory,'' Class.\ Quant.\ Grav.\ {\bf 16}, L37 (1999) [arXiv:hep-th/9807241].} for a similar discussion. }. So it seems most probable that in quantum
gravity the wavefunction for these closed universes is completely determined. To the extent that the field theory
on the boundary determines the $\alpha$ parameters, then the wavefunction for the closed universes are also
determined by the field theory . It would be nice to understand to what extent the field theory contains some
information about these closed universes\foot{ Hopefully, the techniques of \ref\FidkowskiNF{ L.~Fidkowski,
V.~Hubeny, M.~Kleban and S.~Shenker, ``The black hole singularity in AdS/CFT,'' arXiv:hep-th/0306170.
%%CITATION = HEP-TH 0306170;%%
}
 could be extended to probe
the singularities of the Lorentzian closed universes.}.

On the other hand, it might be that geometries  with
two  boundaries correspond to a partition function of
the form $Z = \sum_i Z_{i}^{1} Z^2_i $ where the index $i$ runs over
some ``sector'' of the field theory and the indices $1,2$ indicate
the two field theories at each boundary. Such a situation might
arise if the  field theory
 partition functions are not well defined\foot{
The boundary field theories discussed in sections 4,5 seem perfectly
well defined in this sense.
In the (0,4) 2d CFT
associated with the $AdS^3\times S^3/Z_N \times K3$ example in section
3 there could be subtleties due to the fact that the theory is not
left/right symmetric.  }. An example of such a
theory is a chiral boson in two
dimensions. One way  to define a modular invariant answer
is to consider a left and right moving sector. It
was observed in  \ref\WittenWY{ E.~Witten, ``AdS/CFT correspondence and topological field theory,'' JHEP {\bf
9812}, 012 (1998) [arXiv:hep-th/9812012].
%%CITATION = HEP-TH 9812012;%%
} that the $AdS_5 \times S^5$ partition function has a similar property.
On the other hand the effects discovered
in \WittenWY\ are related to the overall U(1) degree of freedom.
So by adding a simple U(1) field at the boundary one can get a well
defined partition function\foot{
 If one
imposes local boundary conditions for the
NS and RR two form potentials, then one recovers this $U(1)$ degree of
freedom \ref\MaldacenaSS{
J.~M.~Maldacena, G.~W.~Moore and N.~Seiberg,
%``D-brane charges in five-brane backgrounds,''
JHEP {\bf 0110}, 005 (2001)
[arXiv:hep-th/0108152].
%%CITATION = HEP-TH 0108152;%%
}.}.

In summary, a better understanding of the physics associated to these
geometries is  needed. This might lead to interesting insights
on how quantum gravity works.

{\bf Acknowledgements}

We would like to thank S. Cherkis, R. Dijkgraaf, O. Lunin, J. Polchinski, S-J Rey, K. Skenderis, A. Strominger
 and E. Witten for discussions.
This work was supported in part by DOE grant DE-FG02-90ER40542. LM would like to thank Stichting FOM for support.

\appendix{A}{ Details about solutions involving gauge fields.}

\subsec{ Generalities about merons and instantons}

Let us consider an SU(2) group element $g= e^{{i\over 2}\phi\sigma^3} e^{{i\over 2}\theta\sigma^1}e^{{i\over
2}\psi\sigma^3}$, this parameterizes an $S^3$ \eqn\metrsth{ ds^2_{S^3}=-{1\over 2}Tr(g^{-1}dgg^{-1}dg) = {1\over
4}[( d \psi + \cos\theta d\phi)^2 + d\theta^2 + \sin^2\theta d\phi^2 ] = { 1 \over 4} w^a w^a} Where we  defined
the left invariant one forms $w^a$ through
 $ i { \sigma^a \over 2} w^a = g^{-1} d g $.
We define the SU(2) gauge field $A = i { \sigma^a \over 2} A^a $. The field strength is $F = dA + A^2$ or $F^a =
dA^a - { 1\over 2} \epsilon^{abc} A^b \wedge A^c$. We will be interested in solutions where $A = f(r) g^{-1} dg $,
or $A^a = f w^a$. If $f=1$ then we have a pure gauge configuration. The field strength is $F = f' dr \wedge g^{-1}
dg + f(f-1) g^{-1} dg \wedge g^{-1} dg$. In four flat dimensions we find that the self duality condition implies
$f^{-1} = 1 + { r_0^2 \over r^2}$. In order to find the self dual configuration in $AdS_4$ we start with  the
metric as $ds^2 = d\rho^2 + \sinh^2 \rho ds^2_{S^3}$. We see that defining $ dr/r = d\rho/\sinh\rho$, ie
\eqn\randrho{ r = \tanh \rho/2 }
 we can make the $AdS$ metric conformally flat. This implies that the
 self dual configuration has $f^{-1} = 1 + C/\tanh^2\rho/2$.
 Note that as $\rho$ runs from zero to infinity $r$ goes from zero to one.
 This implies that $AdS_4$ is conformal to the unit ball in $R^4$.
If we are interested in instantons on $S^4$ all we need to do is
to change $\rho \to i \gamma$, $ 0 \leq \gamma \leq \pi/2 $ in the
above expressions. Then the function $f$ becomes $f^{-1} = 1 +
C'/\tan^2\gamma/2 $ and for $C'=1$ we find that $f =
\sin^2\gamma/2$. In order to get anti-self dual configurations all
we just change $f \to 1-f$.

For a general metric \eqn\genme{ ds^2 = d\rho^2 + e^{2 \omega(\rho)} ds_{S^3}^2 } the effective action for $f$
becomes \eqn\actionforf{\eqalign{ S &=  \alpha\int F^a_{\mu\nu} F^{a \, \mu\nu}   = 24\alpha\Omega_3 \int d\rho
e^{-\omega}
  [ e^{2 \omega} ( \partial_\rho f)^2  + 4 f^2(1-f^2 ) ]= \cr
 &= 24\alpha\Omega_3\int dy  [(\partial_y f)^2 + 4 f^2(1-f^2)]
 }}
where $dy = d\rho e^{ - \omega(\rho)}$, and where $\Omega_3$ - the volume of a three sphere.

We are interested in evaluating this action for the following three
cases. First when \genme\ is $AdS_4$ and the solution for $f$ is that
of a self dual configuration
 such that the value of $f$ at the boundary of $AdS_4$ is
$f_B$. Then we find that \actionforf\ is \eqn\actioninst{ S_{sd} = 16\alpha\Omega_3 f_B^2 ( 3 - 2 f_B) =
8\alpha\Omega_3[ 1 + 3(f_B- { 1 \over 2}) -4(f_B -1/2)^3] } For an anti-self dual configuration
 with the same boundary conditions we find
\eqn\actionanti{ S_{anti-sd} = 16\alpha\Omega_3[ - f_B^2 ( 3 - 2 f_B) + 1] } Of course we can get \actionanti\
from \actioninst\ by replacing $f_B \to 1-f_B$. For $f_B=1/2$ both have the same action, but for $f_B<1/2$ the
self-dual configuration has lower action than the anti-self-dual configuration. Note that the instanton number
inside $AdS$ is less than one. For $f_B=1/2$ it is $1/2$.

In the case that we have an instanton on $S^4$ the action is the
usual $S_{inst} =  \alpha \int F^a_{\mu\nu} F^{a \, \mu\nu}  = 16
\alpha   \Omega_3 = 32 \pi^2 \alpha $.

\subsec{ Four dimensional solutions}

The renormalized gravity action is \eqn\gravads{\eqalign{ S_{GR} = & \int d^4x \sqrt{g}[-(R+6) ]-2\int
d^3x\sqrt{\gamma} K +(4\int d^3x\sqrt{\gamma} +\int d^3x \sqrt{\gamma}R_{\gamma}) }} where $\gamma$ is the metric
at the regularized boundary, $R_\gamma$ is its curvature and $K$ is the extrinsic curvature of the regularized
surface \ref\Kostasbdy{M.~Henningson and K.~Skenderis,
``The holographic Weyl anomaly,'' JHEP {\bf 9807}, 023 (1998)
[arXiv:hep-th/9806087].}. If the metric is
$AdS_4$, ie. of the form \genme\ with $e^\omega = \sinh(\rho)$ we find that \gravads\ is equal to $S_{GR} = 4
\Omega_3 $, when we take away the regulator. So the total action for the one boundary solution is \eqn\acsing{
S_{1~bdy} = (4 + 8 \alpha ) \Omega_3} After multiplying by the overall normalization factor of the action we get
\actoneb .

Now we consider the case where $f = 1/2$ which leads to the two boundary
solution. For a metric of the form \gravads\
 one can write the gravity action \gravads\ plus the Yang-Mills
contribution as
\eqn\regact{
S \sim  6 \int^{\rho_M} d\rho ( - {\omega'}^2 - e^{- 2 \omega} - 1 + \alpha e^{-4
\omega}  ) e^{3 \omega}
 + 4 e^{ 3 \omega_M} + 6 e^{\omega_M}
}
where $\alpha$ is the coefficient of the gauge kinetic term and $w_M\equiv w(\rho_M)$.
As usual, we should look for a zero energy solution of this action, which
implies
\eqn\eommer{
{\omega'}^2 - e^{- 2 \omega} - 1 + \alpha e^{-4
\omega} =0
}
Solving this equation we find \soleinst .
We can now define a new variable $x = e^{\omega}$, use \eommer\ and write
the counter terms in \regact\ as total derivatives to find that
the two boundary action is
\eqn\twobdya{\eqalign{
S_{2~ bdy} = & 12 \left[ \int_{x_0}^\infty dx ( - 2  \sqrt{ x^4 + x^2 - \alpha  } + 2
x^2 +1 )   + { 2 \over 3} x_0^3 + x_0 \right]\Omega_3
\cr
=& ( 30.296..) \Omega_3  ~~~~~{ \rm for} ~~\alpha =1
}}
where $x_0$ is the value of $x$ at $\rho=0$,
\eqn\valxze{
x_0^2 =  \sqrt{ \alpha + {1 \over 4} }  -1/2
}

It is now interesting to consider the action difference $S_{2~bdy} - 2
S_{1~bdy}$. We find that for the example coming from M-theory, when
$\alpha=1$,
 this
difference is positive. But for sufficiently large $\alpha$ ( $\alpha >  3.775 \pm .005 $) this becomes negative.
In such theories the two boundary solution has less action. It is not clear if one can find an example coming from
string theory that has $\alpha$ large enough\foot{ One case we could consider is the squashed $S^7$ of
\ref\AwadaPK{ M.~A.~Awada, M.~J.~Duff and C.~N.~Pope, ``N = 8 Supergravity Breaks Down To N = 1,'' Phys.\ Rev.\
Lett.\  {\bf 50}, 294 (1983).
%%CITATION = PRLTA,50,294;%%
} this leads to a theory in four dimensions that contains SO(3) gauge bosons, but it is not clear whether one can
truncate it to pure 4d gravity plus the SO(3) gauge field. }.

\subsec{Stability of the 2-boundary solution}

We now consider small fluctuations of the two boundary
solution of the form $f = 1/2 + \epsilon(\rho)$ with
$\epsilon \ll 1$. We substitute this in \actionforf\ and we find
\eqn\quadract{
S_{quad} = 24\int_{-y_m}^{y_m} dy [
( \partial_y \epsilon)^{2}-2\epsilon^2 ]
}
where
\eqn\valym{
 y_m = \int_0^\infty  d\rho e^{- w} =
\int_0^\infty d\rho w' e^\omega { e^{- 2 \omega} \over w'} =
 \int_{x_0}^\infty
{ dx \over \sqrt{ x^4 + x^2 - \alpha }}
}
and $x_0 \equiv \sqrt{\alpha+{1\over 4}}-{1\over 2}$. The eigenvalue problem for this solution is very simple, we impose
 that $\epsilon$ vanishes at $y = \pm y_m$.
The lowest mode has the form $\cos ( \pi y/(2 y_m) )$ and
it leads to the  eigenvalue
\eqn\negeig{
\lambda_0=-2+({\pi\over {2 y_m}})^2
}
For $\alpha =1$ this is negative. For $\alpha >  1.245 \pm 0.005$ this
is positive.

\subsec{ The M-theory lift of the solutions}

Using the formulas in \Cveticfour\
 with the two equal gauge fields, $A^a = {\tilde A}^a$,
 we get the 11-dimensional solution
\eqn\soleleven{\eqalign{ ds^2_{11} &= ds_4^2 + 4d\xi^2 +
\cos^2\xi \sum_a (w^a-A^a_\mu dx^\mu)^2 + \sin^2\xi
\sum_a (\tilde{w}^a-A^a_\mu dx^\mu)^2 \cr F_{(4)} &= -3\epsilon_{(4)}+\sqrt{2}\sin\xi\cos\xi d\xi\wedge
(w-\tilde{w})^a\wedge *F_2^a +\cr &+{\sqrt{2}\over 4}\epsilon_{abc}
[\cos^2\xi(w-A)^a\wedge
(w-A)^b + \sin^2\xi(\tilde{w}-A)^a\wedge (\tilde{w}-A)^b]  \wedge *F_2^c \cr}}
where $w^a$ , $\tilde{w}^a$ are $SU(2)$ right
 invariant one forms on two 3-spheres $S^3 , \tilde{S}^3$.
This solution does not display the full  $SO(5)$ symmetry that is present
when the two SU(2) gauge fields of \Cveticfour\ are equal.
We can write it in an SO(5) symmetric fashion by thinking about
$S^7$ as an $S^3$ fibration over $S^4$. This can be done by
 writing the metric in \soleleven\ in the following way.
We parameterize $S^3$ in terms of the SU(2) group element $g$ and
$\tilde S^3$ in terms of the SU(2) group element $\tilde g$.
Then we write the right invariant one forms as
${ i  \sigma^a \over 2} w^a_R = gdg^{-1}$ and
\def\tg{ {\tilde g}}
\def\hg{ {\hat g}}
${ i  \sigma^a \over 2 }\tilde{w}^a_R={\tilde g} d{\tilde g}^{-1}$
where $ \sigma^a $ are the Pauli matrices.
We then define  $\hat g =\tg^{-1}g$ with
${ i  \sigma^a \over 2 }{\hat w}^a_R\equiv
 \hat gd\hg^{-1}$, so that $g=\tg \hg$ and
$\tg^{-1}gdg^{-1}\tg =d\tg^{-1}\tg+ \hg d\hg^{-1}$ and $w_R^2 =-2 Tr (gdg^{-1})^2 = (\tilde{w}_L+{\hat w}_R)^2$.
The $S^7$ metric (of radius $2$) is then \eqn\sseven{ ds^2 = 4d\xi^2+\cos^2\xi w_R^2+\sin^2\xi \tilde{w}_R^2 =
[d\zeta^2 + \sin^2\zeta{1\over 4} {\hat w}_R^2]+(\tilde{w}_L+\cos^2 {\zeta\over 2} {\hat w}_R)^2 } where we used
that $\tilde{w}_L^2 = \tilde{w}_R^2 $ and we  defined $\zeta\equiv 2\xi$ , $0\leq\zeta<\pi$. The coordinates
$\zeta$ and $\hg $ parameterize an $S^4$. We see
 that $B^a \equiv \cos^2 {\zeta\over 2}w^a_R$ can be thought of as the
gauge field of an SU(2) instanton on $S^4$ that is $SO(5)$ symmetric, up to ``gauge'' transformations. The gauge
transformations that act on the gauge field $B$ are the right $SU(2)$ transformations acting on the fiber
parameterized by $\tg$.

Now we are ready to  add the $A$ gauge field appearing in the metric in \soleleven . We see that $- 2 Tr(w_R-A)^2 =
(\tilde{w}_L+{\hat w}_R-\tg^{-1}A\tg)^2$,
where hopefully the notation is self explanatory\foot{
The expression $ \tg^{-1}A\tg $ means that we rotate the gauge indices, $a$,
 of
$A^a_\mu$ by the group element $\tg$.}. The metric then
becomes \eqn\metinsts{\eqalign{  ds^2 &= [d\zeta^2 + {1\over 4}\sin^2\zeta {\hat w}_R^2] +(\tilde{w}_L+B)^2\cr &
-2\cos^2\xi Tr(\tilde{w}_L\tg^{-1}A\tg)-2\cos^2\xi Tr({\hat w}_R\tg^{-1}A\tg)+ A^2-2\sin^2\xi Tr(\tilde{w}_RA)=
\cr &= [d\zeta^2 + {1\over 4}\sin^2\zeta {\hat w}_R^2]+(\tilde{w}_L+B)^2+A^2-2Tr(\tilde{w}_L{\tg}^{-1}A\tg )-
2\cos^2\xi Tr(\tg \omega_R\tg^{-1}A) \cr &= [d\zeta^2 + {1\over 4}\sin^2\zeta {\hat w}_R^2]+
(\tilde{w}_L+B-\tg^{-1}A\tg)^2=\cr &= [d\zeta^2 + {1\over 4}\sin^2\zeta {\hat w}_R^2]+(\tilde{w}_R-A+\tg B
\tg^{-1})^2 }} We see that the two gauge fields $B$ and $A$ are associated to the right and left rotations of the
$\tilde S^3$ fiber parameterized by $\tg $. The final  geometry corresponds to fibering the ${\tilde S}^3$ over
$S^4 \times M^4$ with a gauge field which is the sum of the instanton, $B$,  on $S^4$ plus the gauge field, $A$,
on the four dimensional manifold $M^4$.

\subsec{Checking that this solution is not supersymmetric}

A solution that contains a self dual gauge field configuration on $H_4$
looks like a supersymmetric configuration, since instantons are usually
associated to supersymmetric configurations. In this case it is possible
to check that the solution is not supersymmetric.

In order to do this analysis it is useful to consider the supersymmetries
of SO(4) gauged supergravity given in  \refs{\Das,\Gates}.
In this  theory,
apart of the bosonic fields discussed in \Cveticfour\
there are also the following fermionic fields:
four $\psi^i_\mu$ spin 3/2 Majorana spinors with a vector index in $SO(4)$, and 4 spin 1/2 Majorana spinors
$\chi^i$.\foot{
In order to match conventions in \refs{\Das,\Gates} to the ones we used,
which
are the ones in \Cveticfour
we set $\kappa=1$, set
$g_+=2g , g_-=0$ , relate $A,B$ to $\phi,\chi$ by
 $W=-A+iB=e^{i\sigma}\tanh (\lambda/2)$ with
$\cosh\lambda=\cosh\phi+{1\over 2}\chi^2e^\phi$ , $\sinh\lambda\cos\sigma=\sinh\phi-{1\over 2}\chi^2e^\phi$ ,
$\sinh\lambda\sin\sigma=\chi e^\phi$.}

Also we should separate $F^{ij}$ to the two $SU(2)$ fields - we do
this in the usual way: $J^{12}+J^{34}\equiv iL^1$,
$J^{12}-J^{34}\equiv i\tilde{L}^1$ , $J^{13}-J^{24}\equiv iL^2$
etc. (so that $[L^a,L^b]=i\epsilon^{abc}L^c$,same for the tildes
and the $L^a$, $\tilde{L}^a$ commute among themselves). This is
such that $A^{ij}J^{ij}=iA^aL^a+i\tilde{A}^a\tilde{L}^a$, where
$A^a,\tilde{A}^a$ are real.

In the background we are interested in we have $\phi= \chi =0$. Then the (Lorentzian) susy transformations for the
spinors become \eqn\susy{\eqalign{ \delta \bar{\chi}^i &= {{1}\over
{2\sqrt{2}}}\epsilon^{ijkl}\bar{\epsilon}^j\gamma^{\mu\nu}F_{\mu\nu}^{kl} \cr \delta \bar{\psi}^i_\lambda &=
\bar{\epsilon}^i\overleftarrow{D}_\lambda-{i\over
2}\bar{\epsilon}^j\gamma_\lambda\gamma^{\mu\nu}F_{\mu\nu}^{ij}+ig\bar{\epsilon}^i\gamma_\lambda \cr}} where the
covariant derivative is: $D_\lambda \chi^k= ((\partial_\lambda +{1\over
2}\omega_{\lambda,ab}\gamma^{ab})\delta^{kl}+2gA_\lambda^{kl})\chi^l$.
Let us first take the dagger of these  and multiply by $\gamma_0$ from the left:
\eqn\susy{\eqalign{ \delta {\chi}^i &= -{1\over
{2\sqrt{2}}}\epsilon^{ijkl}F_{\mu\nu}^{jk}\gamma^{\mu\nu}\epsilon^l \cr \delta \psi^i_\lambda &=
D_\lambda\epsilon^i-{i\over 2}F_{\mu\nu}^{ij}\gamma^{\mu\nu}\gamma_\lambda\epsilon^j-ig\gamma_\lambda\epsilon^i
\cr}}
Changing from $SO(4)$ to $SU(2)\times SU(2)$ language, and setting the two $SU(2)$ gauge fields to be equal: $A=\tilde{A}$ so that $F^{23}=F^{24}=F^{34}=0$, we can write
the susy equations in the following way, decomposing the $SO(4)$ $\epsilon^{1,2,3,4}$ into the $SU(2)$ :
$\eta,\epsilon^i$ , $i=1,2,3$: \eqn\susykeep{ \eqalign{ 0 &= F_{\mu\nu}^i\gamma^{\mu\nu}\epsilon^i \cr 0 &=
F_{\mu\nu}^i\gamma^{\mu\nu}\eta \cr 0 &= (\nabla_\lambda-ig\gamma_\lambda)\eta \cr 0 &=
(\nabla_\lambda-ig\gamma_\lambda)\epsilon^i
-\varepsilon^{ijk}(F_{\mu\nu}^j\gamma^{\mu\nu}\gamma_\lambda+4igA_\lambda^j)\epsilon^k \cr}}

Now we perform in the standard way a Wick rotation from the Lorentzian to Euclidean signature \PvN. One finds that the
equations retain the same form in Euclidean space.

We see that the equations for $\epsilon$ and $\eta$ in \susykeep\ decouple. Consider first the conditions on
$\eta$: the gravitino variation implies that $\eta$ is a usual $AdS_4$ Killing spinor. Then we need to further
impose that $F^i_{\mu\nu}\gamma^{\mu\nu}\eta=0$. If we have a self-dual field then
$F_{\mu\nu}\gamma^{\mu\nu}={1\over 2}(F_{\mu\nu}+*F_{\mu\nu})\gamma^{\mu\nu}=F_{\mu\nu}\gamma^{\mu\nu}P_L$. It
turns out that this condition implies the chirality condition $\Gamma^5 \eta =\eta$. One can check that this is
not compatible with the conditions for an $AdS_4$ Killing spinor.
Note however that in the limit that the instanton has zero size this
chirality projection condition is imposed only at the location of the zero
size instanton. This is compatible with the Killing spinor equations in
AdS.
So it appears that a zero size instanton in $AdS_4$ is
supersymmetric.

Now we consider supersymmetries generated by $\epsilon$. Before doing this, note that there is a truncation of
SO(4) gauged supergravity to an $SO(3)$ gauged theory with 3 supersymmetries (the ungauged version of this $SO(3)$
theory was considered in \ref\FreedmanNF{ D.~Z.~Freedman, ``SO(3) Invariant Extended Supergravity,'' Phys.\ Rev.\
Lett.\  {\bf 38}, 105 (1977).
%%CITATION = PRLTA,38,105;%%
}). This theory contains the metric,
three
gravitons, three gauge fields, and one fermion. The supersymmetry
variation
of the fermion and gravitino are given by the $\epsilon$ dependent terms in
\susykeep .

First note that as $F_{\mu\nu}\gamma^{\mu\nu}= F_{\mu\nu}\gamma^{\mu\nu}{1\over 2}(1-\Gamma^5)$, the first
equation in \susykeep\ implies that $\gamma^{\rho i}\epsilon^i=0$. If we use notations where $\epsilon^i_\pm\equiv
{1\over 2}(1\mp \Gamma^5)\epsilon^i$ this means: \eqn\condi{\sigma^i\epsilon^i_\pm=0} where $\sigma^i$ are the pauli matrices. Then the last equation in
\susykeep\ becomes \eqn\susypm{\eqalign{ &\partial_\rho\epsilon_-^i-ig\epsilon_+^i =0 \cr
&\partial_\rho\epsilon_+^i -ig\epsilon_-^i=-4ie^{-2w}f(1-f)\epsilon^{ijk}\sigma^j\epsilon_+^k \cr
&D_\alpha^{ij}\epsilon_-^j = -g\sigma_\alpha\epsilon_+^i \cr &D_\alpha^{ij}\epsilon_+^j =
-4ie^{-2w}f(1-f)\varepsilon^{ijk}\sigma^j\sigma_\alpha\epsilon_-^k+g\sigma_\alpha\epsilon^i_-  }} where $\alpha$
is a curved index on the $S^3$,
$$e^{w(\rho)}=\sinh \rho\,,\,f(\rho)={f_B\sinh^2\rho/2 \over \cosh^2\rho/2-f_B}\,,\,A_\alpha^i=f(\rho)w^i_\alpha$$ and $D_\alpha^{ij}\equiv
(\partial_\alpha-{i\over 2}\sigma_\alpha)\delta^{ij}-4ig\epsilon^{ikj}A_\alpha^k$.

We multiply each of the last two equations in \susypm\ by $\sigma^i$,
summing over $i$. Using \condi\ we find the derivative terms vanish.
Multiplying by $w^{\alpha}_m$ and summing over $\alpha$ we are left with the
following set of algebraic equations:
\eqn\newset{\eqalign{ [-{1\over 2}\delta^{im}+i\varepsilon^{imk}(-{1\over
2}+4igf)\sigma^k]\epsilon^i_-+2ig[-{1\over 2}\delta^{im}-{i\over
2}\varepsilon^{imk}\sigma^k]\epsilon^i_+ &=0 \cr [-{1\over 2}
\delta^{im}+\varepsilon^{imk}(-{1\over 2}+4igf)]\epsilon^i_+
-k2ig[-{1\over 2}\delta^{im}-{i\over
2}\varepsilon^{imk}\sigma^k]\epsilon^i_- &=0 }} where $k\equiv 1+{8\over
g}e^{-2w}f(1-f)$.

These are 6 homogenous equations for 6 spinors $\epsilon^i_\pm$. It is easy to verify
that the determinant of coefficients is nonzero, and thus there is no
nonzero solution to this system.

we conclude that there are also no $\epsilon$ type susys and therefore
this background is not supersymmetric.

\subsec{The conformal field theory}

The dual field theory is the field theory on a stack of $N$ coincident $M2$ branes with an external gauge field
coupled to the R-symmetry currents. This gauge field is given by the boundary value of the bulk gauge fields.
Though it is not possible to write the lagrangian for the interacting theory, it is possible to write a lagrangian
for a single $M2$ brane. A single M2 branes is described by a supersymmetric theory with 8 free real scalar fields
$\phi^a$ transforming as the vector of SO(8) and 8 free real fermions $\psi^\alpha$ transforming as the antichiral
spinor representation, ${\bf 8_s}$ of $SO(8)$. The flat space lagrangian is\foot{ Note that in Lorentzian
signature the spinors $\psi$ are real, so $\bar \psi = \psi^t C$. In Euclidean space we define $\bar \psi = \psi^t
C $, where $C$ is such that $\gamma_i^t = C \gamma^i C^{-1}$ and $C^t = -C$.}
 \eqn\flatmtwo{
S_{M2}= \int i\bar{\psi}^\alpha \sp \psi_\alpha +
(\partial\phi^a)^2
}

The susy transformations for this theory are parameterized by a
spinor in the chiral spinor representation, ${\bf 8}_c$
 of SO(8). they are:

\eqn\susytrs{\eqalign{ \delta \phi^a &= \bar{\epsilon}\Gamma^a\psi
\cr \delta \psi &=  i \sp \phi^a\Gamma^a\epsilon}}

Now we need to
put this on
a 3-sphere and add the gauge field.

Putting the theory on the sphere
makes the derivatives become covariant derivatives with the
spin-connection on the sphere. In addition, to preserve conformal symmetry
we need to add an $\phi^2 R$ term to the Lagrangian.
This also adds another term in the supersymmetry transformation laws.
One finds
\eqn\actsph{
S=\int d\Omega_3 [D_\mu\phi D^\mu\phi+i\bar{\psi}\cd\psi+{3\over
4}\phi^2]
}
where we have set the radius of the $S^3$ to one.
The supersymmetry transformations are
\eqn\susypsh{\eqalign{
\delta_\epsilon \phi & = \bar{\epsilon}\psi
\cr
 \delta_\epsilon \psi &  = \cd (\phi\epsilon)+i\phi\epsilon
}}
where $\epsilon$ obeys $D_\mu\epsilon=-{i\over
2}\gamma_\mu\epsilon$.
Now in order to introduce the gauge field all we need to do is to
add the  gauge fields to the  covariant derivatives in \actsph \susypsh ,
$D_\mu \to D_\mu + A_\mu$ where $A_\mu$ is the boundary value of the
$SU(2)$ gauge field we considered.

\subsec{Details on the five dimensional solution}

We start with the gauge fields in \gaguef\ and \gfields .
We then compute
\eqn\compu{
F^{IJ} = dA^{IJ} + [A,A]^{IJ} = i F^a L^{a~IJ} + i {\tilde F}^a {\tilde
L}^{a~IJ}
}
We chose the gauge fields so that $* F^a = F^a$ and $*{\tilde F}^a = -
{\tilde F^a}$.
We take both instantons to be SO(5) spherically symmetric.
Using that $F^a_{\mu \nu} F^{b ~ \mu\nu} = {\tilde F}^a_{\mu \nu}
{\tilde F}^{b ~ \mu\nu} = 4 \delta^{ab}$ and
$ F^a_{\mu \nu} {\tilde F}^{b ~ \mu\nu} =0$  we
 can compute
\eqn\lagrfsq{
F^{IJ}_{\mu\nu} F^{KJ \, \mu \nu} =  - ( L^{a~IJ}L^{a~K J} + {\tilde L}^{a~IJ}
{\tilde L}^{a~KJ}) 4  = 8 \delta^{IK}
}
where we used that $J^{a~KJ} = - J^{a ~JK}$ and that $ (J^a J^a)^{IK} =
2 \delta^{IK}$, $J^a=L^a,
\tilde{L}^a$. Note that this implies that
$ { 1 \over 4} F^{IJ} F^{IJ} = 12$, which in turn gives $\beta =1$ in
\effact , after using the action in \Cveticfive .
We now need to observe that
\eqn\observe{\eqalign{
F^a \wedge F^b =& - {\tilde F}^a \wedge {\tilde F}^b \sim \delta^{ab} \epsilon_4
\cr F^a \wedge {\tilde F}^b =& 0
}}
This implies that
\eqn\wedgedf{
\epsilon_{IJKLMN} F^{KL}\wedge F^{MN} \sim
\epsilon_4 ( L^{a \, KL} L^{a \, MN} - {\tilde L}^{a \, KL} {\tilde L}^{b
\, MN}) =0
}
This implies that we are obeying the Chern Simons equations.

\subsec{ Search for negative modes from gauge fields}

The background fields are given by
\eqn\backgr{\eqalign{
A^{ab} = &  f \epsilon^{abc} w^c~,~~~~~~~~~~~~~
A^{AB} =  {\tilde f} \epsilon^{ABc} w^c
\cr
f = & \cos^2\theta/2~,~~~~~~~~~~~~~~\tilde f = 1-f=\sin^2\theta/2
}}
We can now compute
\eqn\backgr{
F^{ab} =  f' d\theta \epsilon^{abc} w^c + f(1-f) w^aw^b
}
We now consider the small fluctuation
\eqn\smpe{
-A^{Ba} = A^{aB} = h \epsilon^{aBc} w^c + g \delta^{aB} d \theta
}
where $h,g$ are functions of $\theta$.
This leads to the following additional terms in $F$
\eqn\fstr{\eqalign{
\delta F^{aB} =& [ h' + g ({\tilde f} - f)] d\theta \epsilon^{aBc} w^c
+ h (1 - f - {\tilde f})
w^aw^B
\cr
\delta F^{ab} =- & h^2 w^a w^b + 2 h g \epsilon^{abc}d\theta w^c
\cr
\delta F^{AB} = + & h^2 w^A w^B - 2 h g \epsilon^{ABc}d\theta w^c
}}
We now compute the action. In order to do this we need to remember that
the unit normalized vielbein on $S^3$ is given by $e^a= w^a/2$.

We can now compute $F^2$ for the above configuration \backgr \smpe\
\eqn\acetot{\eqalign{
 S= & \int_{S^4} F^{IJ}_{\alpha \beta} F^{IJ \, \alpha \beta}
 \cr
  = & 48 \Omega_3 \int d\theta \sin^3 \theta \left[
{ (f' + 2 hg)^2 + ({\tilde f}' - 2 h g)^2 + 2 (h' + g (\tilde f - f))^2
 \over \sin^2 \theta} +\right.
 \cr
 & ~~~~~~~~
 \left. + 4{ [ -f (1-f) + h^2]^2 + [ - \tilde f ( 1 - \tilde f) + h^2]^2
 +  2 (1-f - {\tilde f})^2 h^2
\over \sin^4\theta}   \right]
}}
We see that the above expression vanishes
for
\eqn\valuesga{
g =1/2~, ~~~~~h = {1 \over 2} \sin \theta
}
In fact \valuesga\ correspond to the values we would obtain if we start
with the pure gauge
configuration $ A^{ab} =  \epsilon^{abc} w^c$, $A^{AB}=0$, $A^{aB}=0$
 and we
do a gauge transformation by the gauge group element $
 e^{ i {\theta \over 2} \Sigma}$
where $\Sigma$ is the $U(1)$ that exchanges $1,2,3$ with $4,5,6$.

We see that by turning on $g$ and $h$ continuously we can start
from the instanton/anti-instanton configuration and go to the pure gauge
configuration \valuesga .
Since varying $h$ and $g$ we can find a path that gets
rid of the original gauge field configuration,
we expect to get a negative
mode for small fluctuations of $h$ and $g$.
In principle we need to consider the most general fluctuation of
the gauge fields in order to find all negative modes.
Since the fluctuations
parametrized by $h,g$ are general enough to provide a path in
field space that gets rid of the instanton/anti-instanton
configuration it is very likely that
 the negative mode we find by considering
small fluctuations of $h,g$ is the most negative mode.

Denoting the small fluctuations of the gauge field $A$ around
the background \backgr\ by $\delta A$ we can expand the Yang-Mills
action to second order. This will schematically lead to
an expression of the form $\int \delta A {\cal O} \delta A$, where
${\cal O}$ is some operator. We are interested in finding eigenvalues
for this operators ${\cal O} \delta A = \lambda \delta A$.
In order to do this we consider the auxiliary action given by
\eqn\auxact{
S^{(2)}_{aux} = \int \delta A {\cal O} \delta A - \lambda (\delta A)^2 }
and we look for solutions of this action that are non-singular.
These will exist only for special values of $\lambda$.
If $\lambda \not =0$ we do not have to worry about gauge fixing.
We can compute the first term in \auxact\ by expanding \acetot\
to second order in $h,g$.
We find
\eqn\actse{
S^{(2)}_{aux} = 48 \Omega_3 \int d\theta \sin \theta
\left[  2 (h' - g \cos\theta)^2 -  4 h^2 - 4 h g \sin \theta
- {\lambda \over 4} ( \sin^2 \theta g^2 + 8 h^2)
\right]
}
In order to find the eigenvalue $\lambda$ we just solve the equations
of motion for $S$. The equation of motion for $g$ is algebraic, so we
replace the resulting value back into the action and we get the
following action for $h$
\eqn\actget{
S^{(2)}_{aux} = 48 \Omega_3\int d\theta \sin \theta
\left[ 2 {h'}^2 - 4 h^2 - 2 \lambda h^2 - 4 {(  h' \cos \theta
+ h \sin \theta )^2 \over 2 \cos^2 \theta - { \lambda \over 4}
\sin^2 \theta }
\right]
}
The equations of motion for $h$ read
\eqn\eqmot{\eqalign{
- 2 &( \sin \theta h')' - ( 4 + 2 \lambda) \sin \theta h
  ~ + \cr
+ &4 \left( { \sin \theta \cos \theta ( \cos \theta h' + \sin
\theta h) \over 2 \cos^2 \theta - { \lambda \over 4} \sin^2 \theta
} \right)' - 4 { \sin^2 \theta ( h' \cos\theta + \sin \theta h )
\over 2 \cos^2 \theta - { \lambda \over 4} \sin^2 \theta } =0 }}
Examining the equation near $\theta \sim 0$ we find that the
regular solution goes as $\theta^2$. We find a similar behavior at
$\theta = \pi$. The value of $ \lambda$ should be chosen so that
\eqmot\ has a nontrivial solution which is regular at $\theta =0,
\pi$. We find a regular solution for $\lambda = -4$ which is $h =
\sin^2 \theta$. We see that this is the lowest eigenvalue. We can
check numerically that lower values of $\lambda$ lead to solutions
for $h$ which do not cross the real axis if they are regular at
one of the two ends.

Now we need to consider the radial dependence.
We choose a gauge $A_\rho =0$. The only terms with $\rho$ derivatives
will come from  $F_{\rho \alpha} = \partial_\rho A_\alpha$,
where $\alpha$  is an index on $S^4$.
So we see that for a given eigenvector with eigenvalue $\lambda$ the
lagrangian will be of the form
\eqn\vallagr{\eqalign{
S^{(2)} &\sim \int d\rho \left[  2  e^{ 2 \omega}
 F^{IJ}_{\rho \alpha} F^{IJ}_{\rho \alpha}
+ \lambda A_\alpha^{IJ} A^{IJ}_\alpha \right]
\cr
S^{(2)} & \sim\int d\rho \left[ 2e^{2 \omega} (\phi')^2 + \lambda  \phi^2 \right]
}}
where $ e^{2 \omega}$ was given in \effact\ and
 $\phi$ is a field which encodes
the $\rho$ dependence of the eigenvector with eigenvalue $\lambda$.
In other words, we consider a gauge field fluctuation
$\delta A = \phi(\rho) \delta A^{(\lambda)}$, where
$\delta A^{(\lambda)}$ is the eigenvector from \auxact.

We are now interested in understanding if the operator that
appears in the last line of \vallagr\ has negative modes. We
 impose the boundary condition $\phi = 0 $ at $\rho
= \pm \infty $.
We now write down the equation of motion for \vallagr\
\eqn\eoml{
\partial_\rho ( 2  e^{2 \omega} \partial_\rho \phi) - \lambda \phi =0
}
The operator in \vallagr\ will have a negative mode if the
solution to \eoml\
with boundary conditions $\phi(\rho =0) = 1$ and $
\phi'(\rho =0) = 0$ changes sign as $\rho \to \infty$.
Setting $\lambda =-4$ and
analyzing the equation numerically we can see that it does
not change sign\foot{
 We can have negative modes if $\lambda < -5.67 \pm 0.03$ }

The final conclusion is that there is no negative mode of the
type we looked for. There  might be a negative mode that has
a more complicated wavefunction on $S^4$.

\subsec{ Analysis of the  scalar fields in the instanton background}

Using \gaguef\ and \gfields\ we find that
\eqn\lagrfsq{
F^{IJ}_{\mu\nu} F^{KL \, \mu \nu} =  - 4 (f
 L^{a~IJ}L^{a~K L} + {\tilde L}^{a~IJ}
{\tilde L}^{a~KL})
}
We can now use that if $I,J,K,L$ are all $1,2,3$ then we have\foot{
Note that $L^{a \, IJ} = -\epsilon^{a IJ}$.}
\eqn\jsq{
L^{a~IJ}L^{a~K L}
= - ( \delta^{IK} \delta^{JL} - \delta^{IL}\delta^{JK} )
}

We now turn to a discussion of the scalar fields. They are given in terms
of a matrix $T_{IJ}$ which has determinant one.
This implies that we can write
\eqn\texp{
T^{-1}_{IJ} = \delta_{IJ} + \phi_{IJ}  + h \delta_{IJ} ~,~~~~~~~
12 h = \phi_{IJ}
\phi_{IJ} + \cdots
}
This ensures that $T$ has determinant one to second order.

We now evaluate the kinetic term and potential using the expressions in
\Cveticfive. We find
\eqn\acscal{
S = \int { 1 \over 4} \partial \phi_{IJ} \partial \phi_{IJ}
- g^2 \phi_{IJ}  \phi_{IJ}
}
To evaluate this we need to compute $T_{IJ}$ from \texp\ , which is
$T_{IJ} = (1 + h) \delta_{IJ} - \phi_{IJ} + \cdots$,
 where the dots include
a quadratic term in $\phi$ which is traceless and does not
contribute to the computation leading to \acscal ). After setting
$g=1$ we see that we get the correct mass, $m^2 = -4$. We are
working in units where the radius of $AdS$ is one.

We need to add the term that comes from the coupling of the
scalars to the gauge fields given by
\eqn\couplgf{
{ 1 \over 4} T^{-1}_{IK}T^{-1}_{JL} F^{IJ} F^{KL}
}
The scalars can be decomposed under $SO(6) \to SO(3) \times SO(3)$ as
$20 \to (5,1) + (1,5) + (3,3) + (1,1)$.
Looking at \texp\ we see that there are two kinds of couplings that
can appear from \couplgf , either one  $\phi$ from each from each
$T$ in \couplgf\ or a coupling to $h$  in \texp\ from one of the $T$s
in \couplgf .
 The coupling from $h$ does not depend on the type of $\phi$ we
consider  and is equal to
\eqn\hcoupl{
 { 2 h} { 1\over 4} F^{IJ} F^{IJ} = 2 h 12 = 2\phi_{IJ} \phi_{IJ}
 }
The two $\phi$ coupling will depend on the type of scalar we consider.
For the $(3,3)$ scalars there is no extra coupling and \hcoupl\ is the
full answer.
For the (5,1) scalars we can now use \lagrfsq\ and \jsq\ to find
the extra term
\eqn\extraff{
 -  \phi_{IJ} \phi_{IJ} ~,~~~~~~~~I,J = 1,2,3
 }
We can do a similar analysis for the $(1,1) $ field, which gives us the
extra term
 \eqn\extraoo{
  2 \phi_{IJ} \phi_{IJ}
}
where we used that $\phi$ is diagonal and rewrote it as in \extraoo .

Now we study the question of whether there are negative modes.
We need to consider the equation
\eqn\equat{
 - e^{- 4 \omega} \partial_\rho ( e^{ 4 \omega} \partial_
 \rho \phi)
+ (m^2 + \gamma e^{-4 \omega}) \phi = \lambda \phi
}
where $m^2 = -4$ and  $\omega $ is given in \metricf .
The parameter $\gamma$ is a number that depends on the type
of scalar field. For the scalars in the (3,3),(1,1),(1,5),(5,1)
representations $\gamma = 8, 16,4,4$ respectively.
 It is important to note that $\gamma >0$ for all the
scalar fields considered.

Solving the equation \equat\
with the initial condition $\phi(0)= 1$, $\phi'(0)=0$ we see that
if $\gamma >0$ then the solution stays positive. This means that
the corresponding operator does not have any negative modes.

\listrefs
\bye